\documentclass[12pt]{iopart}

\usepackage{graphicx}
\usepackage{xcolor}

\begin{document}



\title{Interaction of driven ``cold" electron
plasma wave with thermal bulk mediated by spatial ion inhomogeneity}


\author{Sanjeev Kumar Pandey$^{[1]}$ $^{\dag}$ and Rajaraman Ganesh$^{[2,3]}$}

\address{$^{1}$ India Center for Lab Grown Diamonds (Incent-LGD), Material Science Research Center (MSRC), Department of Physics, Indian Institute of Technology (IIT) Madras, Chennai 600036, India}
\address{$^{2}$ Institute for Plasma Research (IPR), Bhat, Gandhinagar 382428, India}
\address{$^{3}$ Homi Bhabha National Institute (HBNI), Mumbai, Maharashtra 400094, India}
\address{$\dag$ Part of this work was done by one of the Authors (SKP) during his PhD Thesis work at the Institute for Plasma Research, Gandhinagar.}
\ead{sanju23510@gmail.com}
\vspace{10pt}
\begin{indented}
   \item July 2024
\end{indented}

\begin{abstract}
 Using high resolution Vlasov - Poisson simulations, evolution of driven ``cold" electron plasma wave (EPW) in the presence of stationary inhomogeneous background of ions is studied. Mode coupling dynamics between ``cold'' EPW with phase velocity $v_{\phi}$ greater than thermal velocity i.e $v_{\phi} \gg v_{thermal}$ and its inhomogeneity induced sidebands is illustrated as an initial value problem. In driven cases, formation of BGK like phase space structures corresponding to sideband modes due to energy exchange from primary mode to bulk particles via wave-wave and wave-particle interactions leading to particle trapping is demonstrated for inhomogeneous plasma. Qualitative comparison studies between initial value perturbation and driven problem is presented, which examines the relative difference in energy transfer time between the interacting modes. Effect of variation in background ion inhomogeneity amplitude as well as ion inhomogeneity scale length on the driven EPWs is reported. 

\end{abstract}
%
\vspace{2pc}
\submitto{\PS}
%
%
%
\section{Introduction}
\label{Introduction}

Electron plasma waves (EPWs) have been extensively researched since its discovery in 1929 by Tonks and Langmuir \cite{tonks_langmuir1929}. Investigation of the mechanism behind its excitation and dynamical response continues to be a fundamental area of research. Several pioneering works have brought out various key aspects associated with the EPWs \cite{bohm1949,landau1946,kampen1955,bgk1957,dawson1959,oneil1965,kds1969,goldman1970}. In the recent past, numerous studies including kinetic or phase space simulations have been carried out to understand intricate features of the driven dynamics at electron and ion scales in homogeneous collisionless plasmas \cite{gary_tokar_1985,valentini_prl_2010,valentini_prl_2011,Valentini_2011,valentini_2012, pallavi_2016,pallavi_2017,pallavithesis,pallavi_2018}. However, in most realistic scenarios like laboratories, tokamaks, and astrophysical plasmas, equilibria are
inhomogeneous in nature and hence studies involving such inhomogeneities becomes a necessity.

In the above context, in the past, studies have been carried out involving temporal evolution of long wavelength or high phase velocity EPWs in the presence of ﬁnite amplitude ion density modulation in the background. These studies demonstrates energy transfer from high phase velocity ``cold'' EPWs into associated sidebands with much slower phase speeds and then into the particles via efﬁcient off-resonant mode coupling mediated by inhomogeneous background of ions \cite{kruer_kaw_1970,kruer_prl1970,kruer1972}. Additionally, attempts were made to investigate the evolution and energy transfer mechanism associated with EPWs in a bounded and periodic inhomogeneous equilibria \cite{jackson1966,harding1968,dorman1970,bertrand_feix_baumann_1971}. In another work, using warm plasma model \cite{kaw1973} and kinetic model \cite{sanjeev2021,pandey_2022_KAW} with initial value perturbation (IVP) approach, authors illustrated damping of high phase velocity EPWs via wave-wave mode coupling phenomenon initiated by inhomogeneous stationary background of ions. This study \cite{kaw1973} showed that an EPW of mode number $k$ can interact with a non-uniform (i.e, spatially periodic), stationary ion background of scale $k_{0}$ to produce coupled secondary sideband modes with wavenumber $|k \pm Nk_{0}|$ (where $N$ is a dimensionless coupling parameter), resulting into cascading energy transfer from primary mode to sidebands upto $N^{th}$ sideband mode. In the process, depending on the value of $N$, the primary perturbation mode may lose its amplitude either fully or partially. These studies are proven useful in various applications and help in understanding variety of plasma concepts such as wave breaking in inhomogeneous equilibria \cite{xu2019,nidhi2021}, laser plasma interaction \cite{everett1995,everett1996} , laser absorption by ion acoustic turbulence \cite{estabrook1981}, plasma instabilities \cite{rosen1972,canosa1976,shoucri1978,Buchelnikova_1980,Buchelnikova_1981,koch1983,barr1986,villeneuve1987,ghizzo1988,brunetti2000,manfredi2000,shoucri_2006,shukla_2009,yang2020,Pandey_2021_TPI_2}, ionospheric turbulence \cite{pottelette1984,guzdar1996} etc.

In the present work, using high resolution OpenMP based kinetic (VPPM-OMP 1.0) Vlasov - Poisson solver with constant frequency external electric field drive, we address the energy exchange dynamics associated in both wave-wave mode coupling as well as wave-particle interaction process. In this process, ``cold" primary perturbation is externally driven in the presence of background ion inhomogeneity which in turn facilitates onset of coupled secondary sidebands of  the externally driven ``cold" EPWs. At first, we demonstrate qualitative agreement between the electron plasma waves excited via initial value perturbation (IVP) and externally driven perturbation (DP) approaches. It is found that despite having identical coupling parameter $N$, in IVP cases some higher coupled sideband modes are involved in energy exchange interactions compared to the DP cases. Also, the exact values of the time required for the transfer of most of energy density from primary driven perturbation mode to the coupled sideband modes differ slightly in both the IVP and DP cases. Next, we illustrate the effect of increase in ion inhomogeneity amplitude (A) and decrease in the ion inhomogeneity scale ($k_{0}$) on the driven dynamics of EPWs, which indicates relatively quick energy transfer between the interacting modes and generation of phase space structures corresponding to a coupled single sideband mode via wave-particle interaction is reported.      

 This paper is organized as follows: In Sec. \ref{Mathematical_model}, we present the Vlasov - Poisson model equations including the equations for stationary ion inhomogeneity and external drive. Followed by numerical scheme in Sec. \ref{Numerical_Scheme}. In Sec. \ref{Simulation_Results}, we present the simulation results for driven EPW cases with different inhomogeneity amplitudes i.e $A=0.03$ (in Sec. \ref{DIP_A=0.03_CASE}) and $A=0.05$ (in Sec. \ref{DIP_A=0.05_CASE}) and finally we conclude in Sec. \ref{Discussion_conclusion}.
  

\section{Mathematical model}
\label{Mathematical_model}

 In the kinetic theory framework, small amplitude externally driven electron plasma waves (EPWs) in a 1D, collision-less, unmagnetized, spatially inhomogeneous plasma system consisting of kinetic electrons and ions is modeled using a set of coupled 1D Vlasov-Poisson (VP) equations \cite{sanjeev2021,pandey_2022_KAW,manfredi1997,raghunathan2013,Saini2018,Pandey_2021_TPI_1,Pandey_2021_TPI_2,sanjeevthesis},
\begin{equation}
\frac{\partial f_{e}}{\partial t}+v_{e}\frac{\partial f_{e}}{\partial x}-E_{T}\frac{\partial f_{e}}{\partial v_{e}}=0
\label{DIP_EQ_1}
\end{equation}
\begin{equation}
\frac{\partial f_{i}}{\partial t}+v_{i}\frac{\partial f_{i}}{\partial x}+ \left[ \frac{E_{T}}{m_{r}} \right] \frac{\partial f_{i}}{\partial v_{i}}=0
\label{DIP_EQ_2}
\end{equation}
\begin{equation}
\frac{\partial E_{T}}{\partial x}= \int f_{i}dv_{i} - \int f_{e}dv_{e}
\label{DIP_EQ_3}
\end{equation}
where $f_{i}(x,v,t) $ and $f_{e}(x,v,t) $ is the ion and electron distribution functions respectively, $m_{r}$ is the mass ratio of ions to electrons, i.e, $m_{r} = M_{i} /M_{e}$, $E_{T}$ is the total electric field given as, 
\begin{equation}
E_{T}=E_{0}+E_{s}+E_{D}^{Ext}
\label{DIP_EQ_4}
\end{equation}
\begin{equation}
E_{D}^{Ext}(x,t)=E_{D}^{0} \sin(kx \pm \omega_{D}^{Lan} t)
\label{DIP_EQ_4_B}
\end{equation}
where $E_{s}$ is the self-consistent electric field, $E_{0}$ is the equillibrium electric field due to background ion non-uniformity and $E_{D}^{Ext}$ is the driven external perturbation electric field to excite EPW, $\omega_{D}^{Lan}$ is the Langmuir driving frequency, $k$ is the mode number and $E_{D}^{0}$ is the amplitude of the external electric field drive. In equation [\ref{DIP_EQ_1}-\ref{DIP_EQ_4}], time $t$ is normalized to electron plasma frequency $\omega_{pe}^{-1}$, spatial coordinate $x$ is normalized to electron Debye length $\lambda_{De}$, velocities to electron thermal velocity $v_{the}=\lambda_{De}\omega_{pe}^{-1}$, electric ﬁeld to $en_{0}\lambda_{De}/\epsilon_{0}$, and distribution function has been normalized to $n_{0}/\lambda_{De}\omega_{pe}$ where $n_{0}$ is uniform plasma density (i.e, $n_{0}$ is the mean density in the absence of inhomogeneity).   

Using Maxwellian velocity distribution function $f_{Me}(v_{e})$ for electrons with an arbitrary spatial inhomogeneity profile $n_{0e}(x)$, the equilibrium solution of Vlasov equation requires (prime denotes spatial differentiation),
\begin{equation}
E_0(x)= - \left[ \frac{n_{0e}^{'}(x)}{n_{0e}(x)} \right]
\label{DIP_EQ_5}
\end{equation}
where $E_{0}(x)$ is the equilibrium electric field resulting from equilibrium inhomogeneity \cite{dorman1970,sanjeev2021}. Ions form a non-uniform, stationary $(\partial/\partial t=0)$ background of number density $n_{0i}(x)$ in this model which is determined by satisfying the equilibrium Poisson's equation,
\begin{equation}
\frac{\partial E_{0}(x)}{\partial x}= \int f_{0i}dv_{i} - \int f_{0e}dv_{e}
\label{DIP_EQ_6}
\end{equation}
Combining Eqs. \ref{DIP_EQ_5}, \ref{DIP_EQ_6} and using equilibrium Maxwellian velocity distribution for electrons and ions alongwith their respective spatial inhomogeneity profiles i.e $n_{0e}(x)$ and $n_{0i}(x)$ one obtains, 
\begin{equation}
\frac{\partial}{\partial x} \left [ - \frac{n_{0e}'(x)}{n_{0e}(x)} \right] = \int n_{0i}(x)f_{Mi}(v_{i}) dv_{i} - \int n_{0e}(x)f_{Me}(v_{e}) dv_{e}
\label{DIP_EQ_7}
\end{equation}
which results in,
\begin{equation}
n_{0i}(x)= n_{0e}(x) + \left[ \frac{(n_{0e}^{'}(x))^{2}-n_{0e}^{''}(x)n_{0e}(x)}{n_{0e}^{2}(x)} \right]
\label{DIP_EQ_8}
\end{equation}
Thus, using Eq. \ref{DIP_EQ_1} - \ref{DIP_EQ_3}, spatial inhomogeneous equilibrium density profile for ions $n_{0i}(x)$ and for electrons $n_{oe}(x)$ are obtained self-consistently, taking into account the equilibrium electric field variation given by Eq. \ref{DIP_EQ_5} . Consequently, one can presume any periodic density profile for $n_{0e}(x)$ and obtain the corresponding $E_{0}(x)$ and $n_{0i}(x)$ from Eq. \ref{DIP_EQ_5} and \ref{DIP_EQ_8} respectively. Note that if $n_{0e}^{''}(x)=n_{0e}^{'}(x)=0$, from Eq. \ref{DIP_EQ_5} and \ref{DIP_EQ_8} we get $E_{0}(x)=0$ and $n_{0i}(x)=n_{0e}(x)$ respectively \cite{sanjeev2021,Pandey_2021_TPI_1,Pandey_2021_TPI_2,pandey_2022_KAW}.

To construct an inhomoegeneous equilibrium, we consider a sinusoidal profile for the spatially inhomogeneous electron density $n_{0e}(x)$ as,
\begin{equation}
n_{0e}(x)= 1+A \sin (k_{0}x)
\label{DIP_EQ_9}
\end{equation}
Using Eq. \ref{DIP_EQ_5} , \ref{DIP_EQ_8} and \ref{DIP_EQ_9} one obtains $n_{0i}(x)$, $E_{0}(x)$ as,
\begin{equation}
E_{0}(x)=- \left[ \frac{(Ak_{0}\cos(k_{0}x))}{(1+A\sin(k_{0}x))} \right]
\label{DIP_EQ_10}
\end{equation}
\begin{equation}
n_{0i}(x)= \left[ \frac{(1+A \sin (k_{0}x))^{3}+A^{2}k_{0}^{2}+Ak_{0}^{2}\sin (k_{0}x) }{1+A^{2}\sin^{2}(k_{0}x)+2A\sin(k_{0}x)} \right]
\label{DIP_EQ_11}
\end{equation} 
where $A$ is the strength of equilibrium inhomogeneity and $k_{0}$ is the measure of equilibrium inhomogeneity scale. To enforce periodicity in the system, which is essential and necessary condition for consistently addressing periodic non-uniform profiles, every equilibrium inhomogeneity scale $k_{0}$ and all the perturbation $k$ scales in the simulation are expressed as integer multiples of $k_{min}$ where $k_{min}=2\pi/L_{max}$. In the following, we present numerical scheme used for solving the coupled Vlasov - Poisson equations.

\section{Numerical Scheme}
\label{Numerical_Scheme}

An in-house developed OpenMP based Vlasov - Poisson solver i.e VPPM - OMP 1.0 \cite{sanjeev2021,Pandey_2021_TPI_1,Pandey_2021_TPI_2,pandey_2022_KAW,sanjeevthesis} is used to solve the set of Vlasov - Poisson equations Eq. \ref{DIP_EQ_1} - \ref{DIP_EQ_3}. VPPM - OMP 1.0 is capable of handling both electron and ions dynamics simultaneously. It is based on Piecewise Parabolic Method (PPM) advection scheme proposed by Colella and Woodward \cite{colella1984} and uses time stepping method given by Cheng and Knorr \cite{cheng1976}. A Fourier transform (FT) based method has been implemented for solving Poisson equation.

 Simulation domain in phase space is chosen to be $D=[0,L_{max}] \times [-v_{e}^{max},v_{e}^{max}]$, where $L_{max}=2\pi/k_{min}$ is the system size and $v_{e}^{max}$ is the maximum electron velocity chosen to be $v_{e}^{max}=6.0$. Periodic boundary conditions (PBC) have been implemented in both spatial and velocity domains and the simulation domain is discretized into $N_{x}$ grid points in spatial domain and $N_{v}$ grid points in velocity domain for both ions and electrons.
 
At $t=0$, we initialize the simulation with normalized Maxwellian distribution function for ions and electrons with respective spatial inhomogeneous equilibrium density profiles $n_{0i}(x)$, $n_{0e}(x)$ obtained earlier (In Eqs. \ref{DIP_EQ_9} and \ref{DIP_EQ_11}),
\begin{equation}
f_{e}(x,v_{e},t=0)= n_{0e}(x)f_{Me}(v_{e})
\label{DIP_EQ_12}
\end{equation}
\begin{equation}
f_{i}(x,v_{i},t=0)= n_{0i}(x)f_{Mi}(v_{i})
\label{DIP_EQ_13}
\end{equation}
\begin{equation}
f_{Me}(v_{e})= \frac{1}{\sqrt{2\pi}} exp \left[ \frac{-v_{e}^{2}}{2} \right]
\label{DIP_EQ_14}
\end{equation}
\begin{equation}
f_{Mi}(v_{i})= \frac{1}{\sqrt{2\pi}}\sqrt{\frac{m_{r}}{T_{r}}} exp \left[ \frac{-m_{r}v_{i}^{2}}{2T_{r}}\right] 
\label{DIP_EQ_15}
\end{equation}
where $f_{Mi}(v_{i})$, $f_{Me}(v_{e})$ are the normalized ion and electron Maxwellian velocity distribution functions, $T_{r}=T_{i}/T_{e}$ is temperature ratio of ions to electrons, $ m_{r}=M_{i}/M_{e}$ is mass ratio of ions to electrons. In the following, we present simulations results for driven EPW cases in the presence of immobile non - uniform background of ions to demonstrate off - resonant wave - wave mode coupling dynamics. Also, we illustrate the comparative study between both IVP and DP cases for identical parameter sets with inhomogeneity amplitude A as A=0.03. 

\section{Simulation Results}
\label{Simulation_Results}
 In this section, we present the simulation results of driven perturbation (DP) studies with $A=0.03$ and qualitatively compare it with small amplitude initial value perturbation (IVP) studies presented in \cite{pandey_2022_KAW}, indicating the existence of coupling of externally driven ``cold'' electron plasma wave (EPWs) with phase velocity $v_{\phi}>> v_{thermal} : v_{\phi} >> v_{e}^{max}$ [where $v_{\phi}$ is the phase velocity of the EPW, $v_{thermal}$ is the thermal velocity of electrons and $v_{e}^{max}$ is the maximum sampled electron velocity] due to the presence of background immobile ion non - uniformity irrespective of chosen perturbation approaches. Also, we present the $A=0.05$ case, to observe the effect of increase in the inhomogeneity amplitude as well as the effect of change in inhomogeneity scale $k_{0}$ on the plasma response of DP cases.

\subsection{Spatially non-uniform driven plasma case with $A=0.03$}
\label{DIP_A=0.03_CASE}

\begin{figure}
\centerline{\includegraphics[scale=0.27]{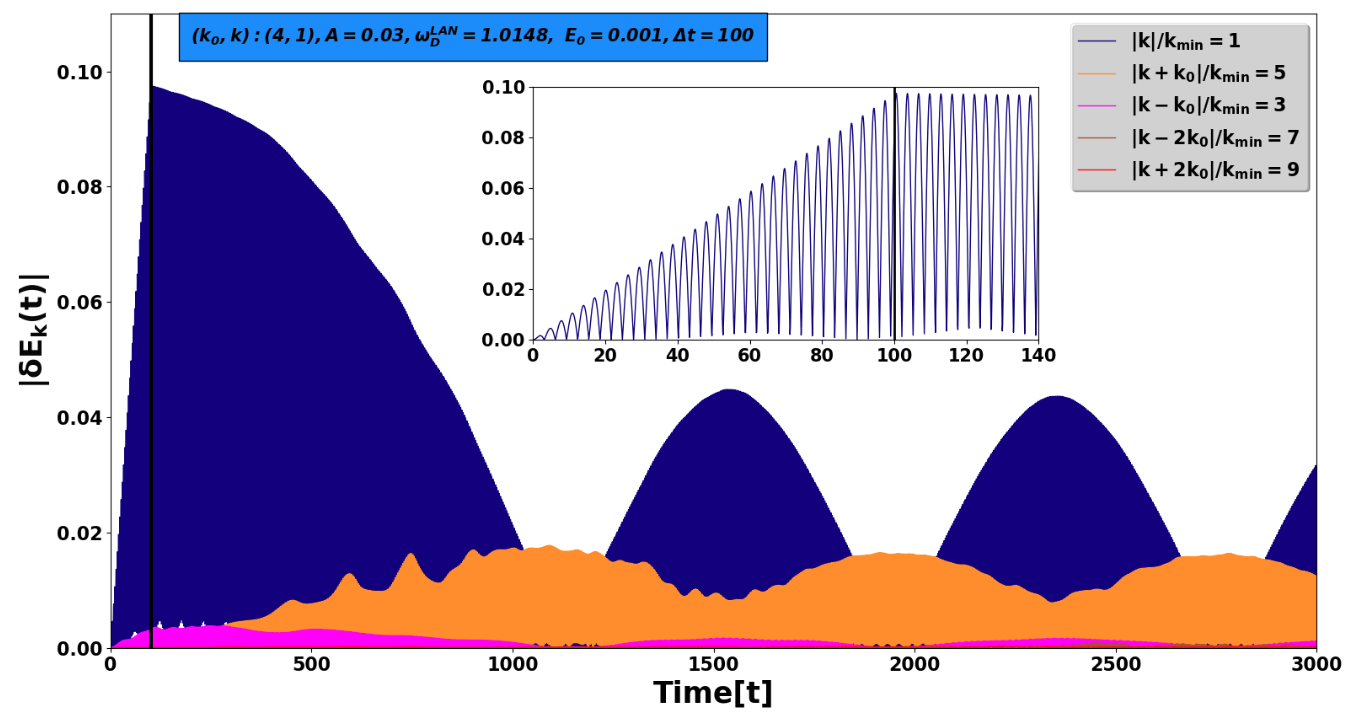}}
\caption{Temporal evolution of primary perturbation $k/k_{min}=1$ mode and coupled secondary interacting $|k\pm Nk_{min}|/k_{min}=3,5,7,9$ sideband Fourier modes i.e $|\delta E_{k}|$ defined by Eq. \ref{DIP_EQ_16} with $(k_{0},k):(4,1)$, $k_{min}=0.1$, $A=0.03$, $\omega_{D}^{Lan}=1.0148$ ($\omega_{D}^{Lan}= \sqrt{1+3k^{2}}$), $E_{D}^{0}=0.001$ and $\Delta t=100$ case. Solid black vertical line indicates the time when the external electric field drive is switched off. Inset plot zooms $|\delta E_{k}|$ for initial time $0$ to $100~\omega_{pe}^{-1}$ indicating the continuous increase in the electric field amplitude value from $0$ to $0.10$ during the external drive.}
\label{41_LNEK_A=0_03}
\end{figure}

As indicated earlier, numerical simulations are carried out in the limit of immobile inhomogeneous background of periodic ions and kinetic electrons. In this plasma system, two spatial scales i.e equilibrium inhomogeneity and perturbation $(k_{0},k)$ scales respectively, are initiated and expressed as an integer multiples of $k_{min}$ which is 0.1 such that $L_{max}=2\pi/k_{min}=20\pi$. To begin with, external electric field drive of the form $E_{D}^{Ext}=E_{D}^{0}sin (kx \pm \omega_{D}^{Lan}t)$ is applied for a time period of $\Delta t=100 ~ \omega_{pe}^{-1}$ with the other simulation parameter set as $v_{e}^{max}=6.0,~(k_{0},k):(4,1),~A=0.03,~\omega_{[D,k_{min}=0.1]}^{Lan}=\sqrt{1+3(0.1)^{2}}=1.0148,~E_{D}^{0}=0.001$. Phase space grid discretization are set to $N_{x} \times N_{v}=[2048 \times 10000]$. To avoid any recurrence effects in the solutions, we chose velocity grid descretization large enough such that $t < T_{R}$, where $T_{R}$ is the recurrence time expressed as $T_{R}=2\pi /k \Delta v : \Delta v = 2 v_{max}/N_{v}$ which is $T_{R}=10 471.98$ \cite{sanjeev2021,raghunathan2013,Pandey_2021_TPI_1,sanjeevthesis,arber_2002,vann_thesis}. The simulation is advanced till $t=3000~\omega_{pe}^{-1}$ in time. Evolution of Fourier mode amplitude of electric field perturbation for mode $k: E_{k}(t)$ is shown in Fig. \ref{41_LNEK_A=0_03} obtained from,
\begin{equation}
 E(x,t)~=~ \sum_{k} E_{k}(t) e^{-ikx}
\label{DIP_EQ_16}
\end{equation}  
where $E(x,t)$ is the total electric field obtained by solving Poisson's equation (Eq. \ref{DIP_EQ_3}).

\begin{figure}
\centerline{\includegraphics[scale=0.21]{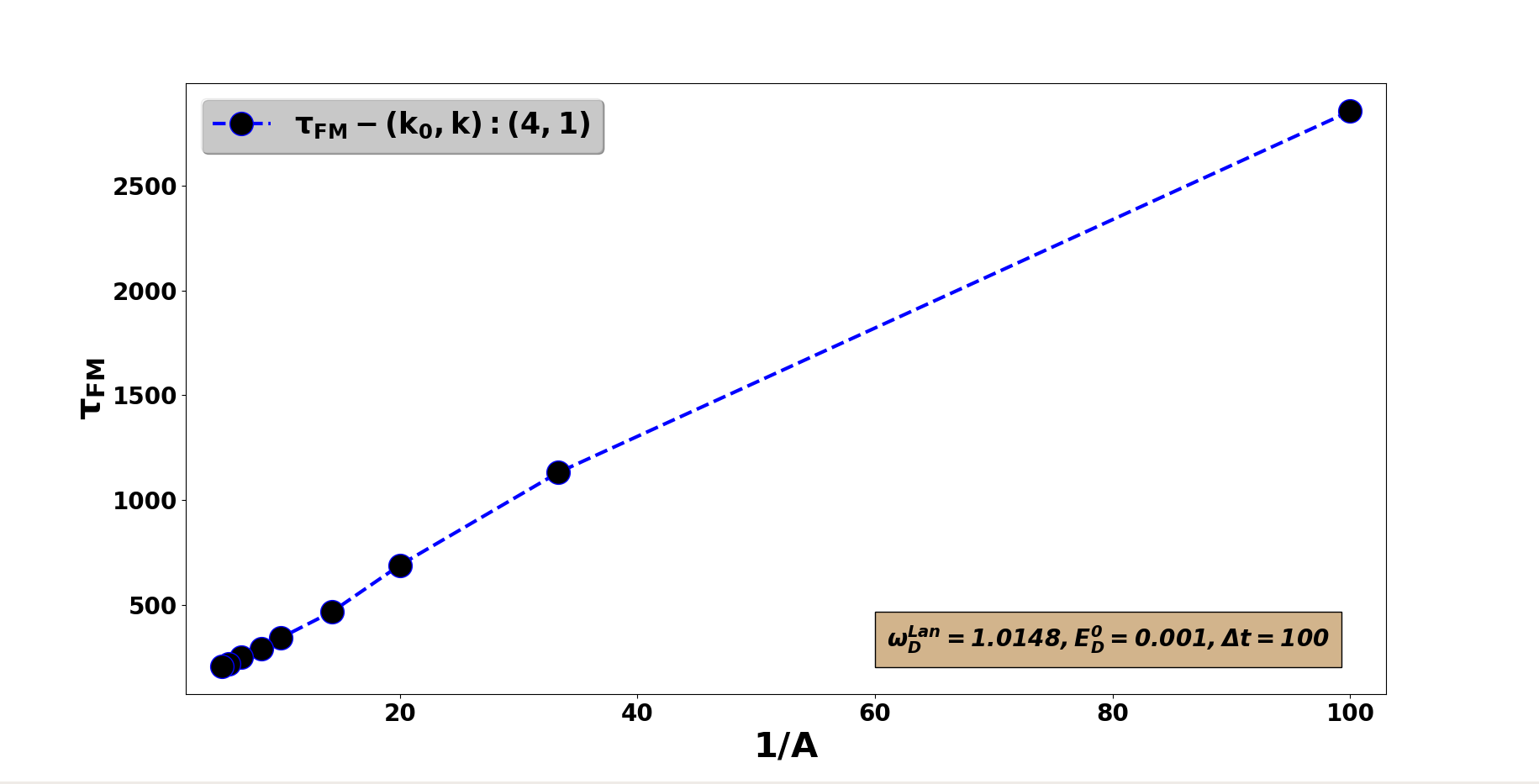}}
\caption{Variation of the $\tau_{FM}$ i.e time taken by the mode to reach the first minimum of the energy density signature versus inhomogeneity amplitude ($A$) for the external driven perturbation case with $(k_{0},k):(4,1)$, $k_{min}=0.1$, $\omega_{D}^{Lan}=1.0148$, $E_{D}^{0}=0.001$ and $\Delta t=100$. Fig. \ref{41_TAU_FM_VS_A} indicates a inversly porprotional relationship between $\tau_{FM}$ and $A$.}
\label{41_TAU_FM_VS_A}
\end{figure}

Fig. \ref{41_LNEK_A=0_03} shows the temporal evolution of the primary $k/k_{min}=1$ and coupled secondary interacting sideband Fourier mode $|\delta E_{k}|$ signatures i.e $k/k_{min}=3,5,7,9$ with $(k_{0},k):(4,1),~k_{min}=0.1,~A=0.03,~\omega_{D}^{Lan}=1.0148,E_{D}^{0}=0.001,\Delta t=100$. It demonstrates the damping of the externally driven primary perturbation mode $|\delta E_{k}|$ signature with phase velocity $v_{\phi}=\omega_{D}^{Lan}/k \simeq 10.148 >> v_{thermal}~: ~v_{\phi}>>v_{e}^{max}$, where $v_{thermal}=1.0$ and $v_{e}^{max}=6.0$, via resonant wave-particle interaction phenomenon associated with the interacting coupled sideband modes $k/k_{min}=3,~5$ in the presence of immobile inhomogeneous background of ions with inhomogeneity amplitude $A=0.03$. For the comparative study, between driven perturbation (DP) and initial value perturbation (IVP) cases, the amplitude of the external electric field drive ($E_{D}^{0}$) and total time for applied external drive ($\Delta t=100$) is chosen such that the total electric field amplitude at the time when external drive is switched off (indicated by solid black line in Fig. \ref{41_LNEK_A=0_03}) is equal to the amplitude of initial electric field [$E_{0}=\alpha/k =0.01/0.1=0.10$] in the initial value perturbation (IVP) case as presented in Sec 3.4 of Ref. \cite{pandey_2022_KAW}. Rest of the parameters in this study are set identical to the initial value perturbation case. On comparing Fig \ref{41_LNEK_A=0_03} of this work with Fig. 19 of Ref. \cite{pandey_2022_KAW} obtained via initial value perturbation approach, one finds a qualitative agreement between both the cases. However, considerable amount of energy transfer in the higher interacting coupled sideband modes $k/k_{min}=7,~9$ can be seen in the IVP cases compared to externally driven (DP) cases despite having the identical mode coupling parameter which is given as $N \sim \sqrt{A/\gamma k_{0}^{2}} \sim \sqrt{0.03/3(0.4)^{2}} \sim 0.25$ \cite{sanjeev2021,pandey_2022_KAW}. 

 In Ref. \cite{pandey_2022_KAW} Sec. 3.2, Authors draw our attention about the time taken by the mode to reach the first minimum of the energy density signature $|\delta E|^{2}$ which is denoted as $\tau_{FM}$. It is the time at which most of the energy content is transferred to the coupled interacting sideband modes from the primary perturbation mode. Fig. \ref{41_TAU_FM_VS_A} illustrates variation of $\tau_{FM}$ versus inhomogeneity amplitude ($A$) for the externally driven perturbation (DP) case with $(k_{0},k):(4,1)$, $k_{min}=0.1$, $\omega_{D}^{Lan}=1.0148$, $E_{D}^{0}=0.001$ and $\Delta t=100$. It indicates that the $\tau_{FM}$ is inversely proportional to the increment in the inhomogeneity amplitude values i.e $\tau_{FM} \sim 1/A$ which is qualitatively similar to the finite inhomogeneity amplitude effects in Ref. \cite{pandey_2022_KAW}. However, there is small difference in the exact values of $\tau_{FM}$ obtained using driven perturbation (DP) and initial value perturbation (IVP)cases i.e $\tau_{FM}^{DP}=1131.91~\omega_{pe}^{-1}$ and  $\tau_{FM}^{IVP}=1099.0~\omega_{pe}^{-1}$ respectively for the identical parameter sets $(k_{0},k):(4,1)$, $k_{min}=0.1$, $A=0.03$. This discrepancy in $\tau_{FM}$ values can be related to the engagement of higher sideband modes $k/k_{min}=7,~9$ during the energy exchange from the primary mode in the IVP case compared to the DP case as indicated earlier.

\begin{figure}
\centerline{\includegraphics[scale=0.27]{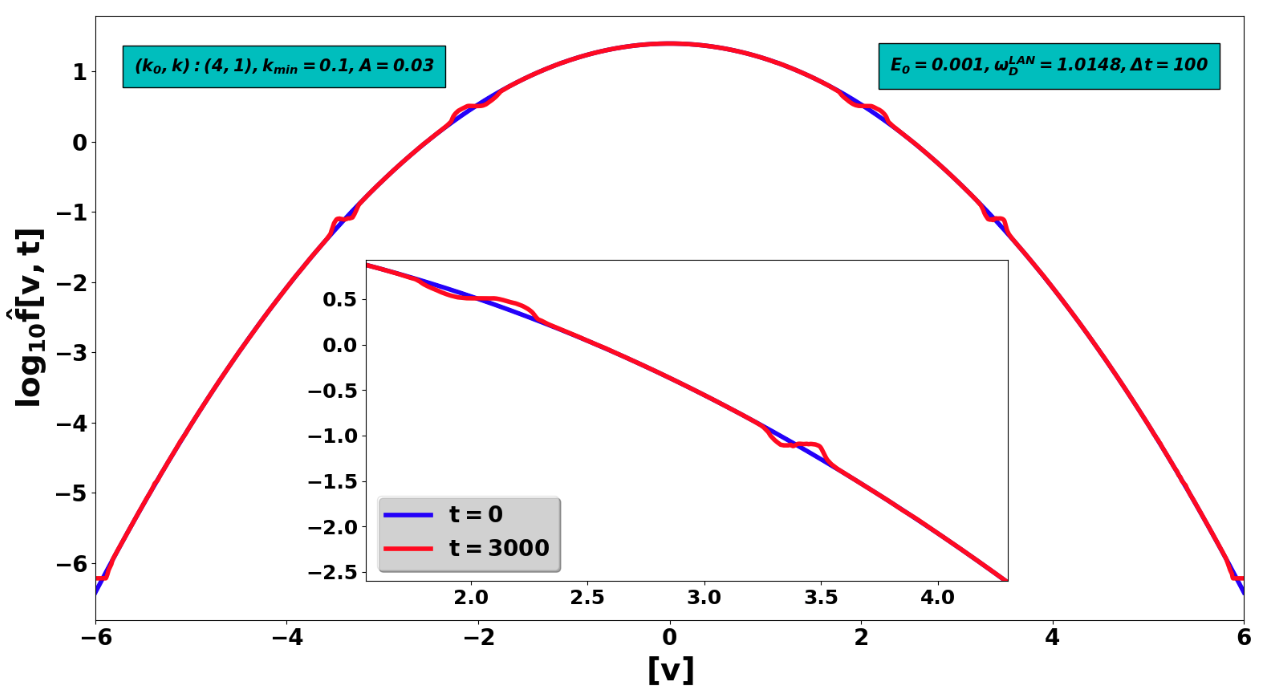}}
\caption{Plot of spatially averaged distribution function $\widehat{f}(v,t)$ with respect to velocity $(v)$ at different times i.e $t=0,~3000 ~\omega_{pe}^{-1}$ for $(k_{0},k):(4,1)$, $A=0.03$, $\omega_{D}^{Lan}=1.0148$, $E_{D}^{0}=0.001$ and $\Delta t=100$ case. In the inset plot, one can observe bump around phase velocity locations $v_{\phi}^{k}=2.04002,~3.38270$ which corresponds to $k/k_{min}=3,~5$ sideband modes. Phase velocities corresponding each interacting mode is tabulated in Table \ref{TABLE_1}.}
\label{41_DFE_A=0_03}
\end{figure}

Resonance locations i.e phase velocities ($v_{\phi}^{k}=\omega_{k}/k$) of each sideband modes is tabulated alongwith oscillation frequency ($\omega_{k}$) obtained by 1D FFT (Fast Fourier Transform) analysis. From Table \ref{TABLE_1}, one can notice that the resonance location corresponding to the primary perturbation mode is ``outside" the velocity domain i.e $v_{\phi}^{k_{min}=0.1}(10.1480) > v_{e}^{max}(6.0)$ which indicates that the wave is ``cold" and that there is absence of resonating wave-particle interactions. However, we have observed wave-particle interaction signatures such as vortex structure formation in the phase space $(x,v)$, plateau creation in the spatially averaged distribution function etc, due to interactions of the secondary sideband modes $k/k_{min}=3,~5,~7,~9$ interacting resonantly with the plasma bulk. 

\begin{table}
\caption[Mode frequency and phase velocity corresponding to the primary and secondary interacting sideband modes ]{Mode frequency $(\omega_{k})$ and Phase Velocity $(v_{\phi}^{k}=\omega_{k}/k)$ corresponding to the primary and secondary interacting sideband modes for $(k_{0},k):(4,1),~E_{D}^{0}=0.001,~A=0.03,~\omega_{D}^{Lan}=1.0148,~\Delta t=100,~k_{min}=0.1$ case.}   
\centering                         
\begin{tabular}{c c c}           
\hline\hline                        
Mode No. & $[\omega_{k}]$ & $[v_{\phi}^{k}=\omega_{k}/k]$  \\ [1.0ex]    
\hline          
$k=0.1$ & 1.01480 & 10.1480 \\          
$|k-k_{0}|=0.3$ & 1.01480 & 3.38270   \\
$|k+k_{0}|=0.5$ & 1.02001 & 2.04002 \\
$|k-2k_{0}|=0.7$ & 1.35509 & 1.93584 \\
$|k+2k_{0}|=0.9$ & 1.01355 & 1.12616 \\ [1ex]
\hline\hline                               
\end{tabular}
\label{TABLE_1}
\end{table}
\begin{figure}
\centerline{\includegraphics[scale=0.30]{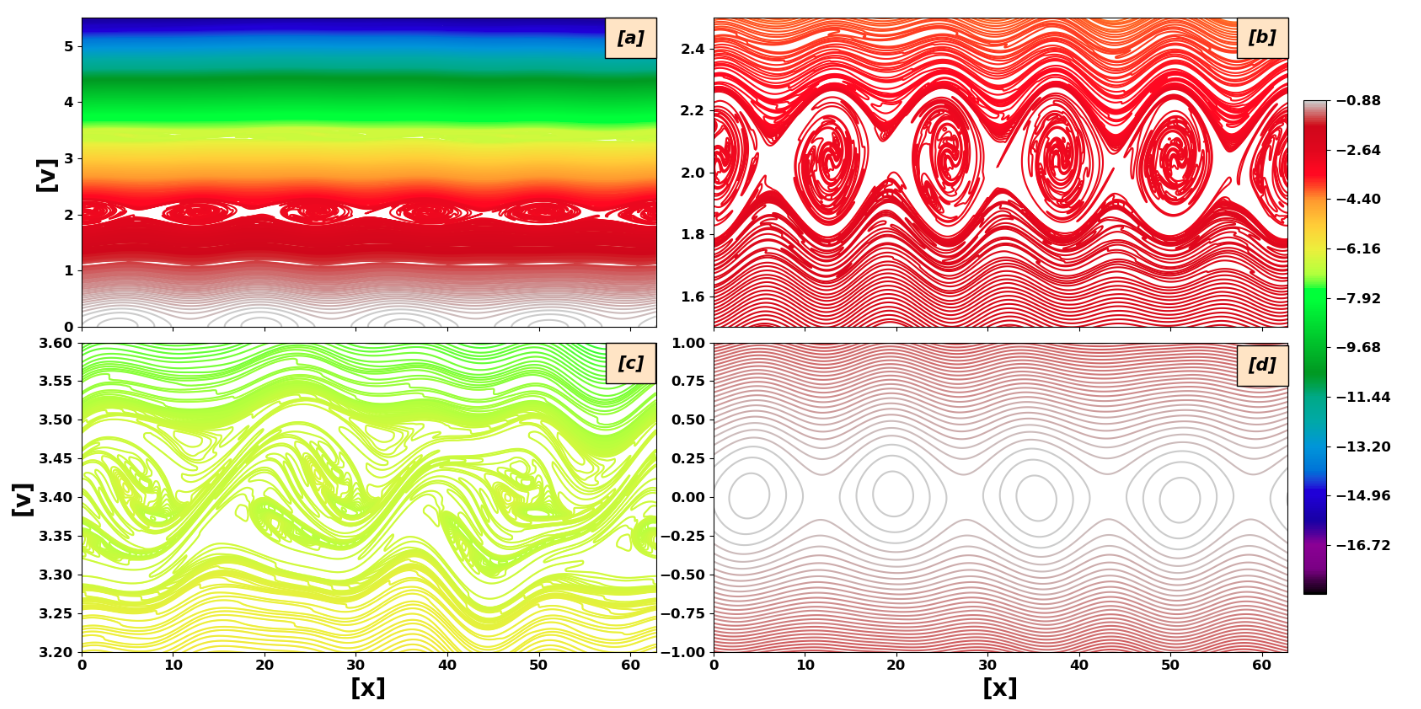}}
\caption{Phase space portrait of electron distribution function $f_{e}(x,v,t)$ at late time $t=3000~\omega_{pe}^{-1}$ for $(k_{0},k):(4,1)$, $A=0.03$, $\omega_{D}^{Lan}=1.0148$, $E_{D}^{0}=0.001$ and $\Delta t=100$ case. In (b) and (c) subplots zoomed velocity ranges to shown to illustrate phase space vortices corresponding to $k/k_{min}=3,~5$ sideband modes around their corresponding phase velocities $v_{\phi}^{k}$ as tabulated in Table \ref{TABLE_1}. In (d) vortex phase space structures around $v=0$ is due to the chosen ion equilibrium inhomogeneity.}
\label{41_CP_A=0_03}
\end{figure}

Spatially averaged distribution function $\widehat{f}(v,t)$ acts as measure to visualize exact state of distribution function around resonance locations and is given as,
\begin{equation}
\widehat{f}(v,t)= \frac{ \int_{0}^{L_{max}} f(x,v,t) dx }{ \int_{ +v_{e}^{max} }^{ -v_{e}^{max} } \int_{0}^{L_{max}} f(x,v,t) dx dv }
\label{DIP_EQ_18}
\end{equation}
where $f(x,v,t)$ is the electron distribution function. Fig. \ref{41_DFE_A=0_03} shows the signature of spatially averaged distribution function with respect to velocity at different times i.e $t=0,~3000 ~\omega_{pe}^{-1}$ with $(k_{0},k):(4,1)$, $A=0.03$, $\omega_{D}^{Lan}=1.0148$, $E_{D}^{0}=0.001$ and $\Delta t=100$. From Fig. \ref{41_DFE_A=0_03} and the inset plot, we can observe formation of plateau or hump around velocity locations $v \sim 2.04,~3.38$ which corresponds to the phase velocity $v_{\phi}^{k}$ locations of sidebands $k/k_{min}=5,~3$ respectively (Table \ref{TABLE_1}). Formation of such plateau or hump structures indicate the wave - particle resonance interaction and particle trapping phenomenon around the phase velocity $v_{\phi}^{k}$ locations as shown in Fig. \ref{41_CP_A=0_03}. Fig. \ref{41_CP_A=0_03} [(a),(b),(c) and (d)] shows phase space plot of electron distribution function at late time $t=3000~\omega_{pe}^{-1}$ for $(k_{0},k):(4,1)$, $A=0.03$, $\omega_{D}^{Lan}=1.0148$, $E_{D}^{0}=0.001$ and $\Delta t=100$ DP case. In a coordinate system moving with the wave speed $v_{\phi}^{k}$, these phase space portrait diagrams illustrates the phase trajectories of the resonant electrons at phase velocity locations $v_{\phi}^{k}$. One can observe in Fig. \ref{41_CP_A=0_03} [(b) and (c)], vortex structure streams corresponding to sideband $k/k_{min}=5$ and $k/k_{min}=3$ respectively. Note that the phase space vortex chain at $v=0$ is due to equilibrium inhomogeneity $k_{0}/k_{min}=4$ with four periodic vortices [Fig. \ref{41_CP_A=0_03} (d)]. It is interesting to note that in the phase space structure of IVP case i.e we observe formation of vortex stream around $v_{\phi}=5.290$ which belongs to $k/k_{min}=2$ sideband mode which is absent in the DP case despite identical parameters as shown in Fig. 20 of Ref. \cite{pandey_2022_KAW} and Fig. \ref{41_CP_A=0_03} of this work. Also, the same is confirmed by the spatially averaged distribution function variations shown in Fig. 21 of Ref. \cite{pandey_2022_KAW} and Fig. \ref{41_DFE_A=0_03}.  
 
\begin{figure}
\centerline{\includegraphics[scale=0.30]{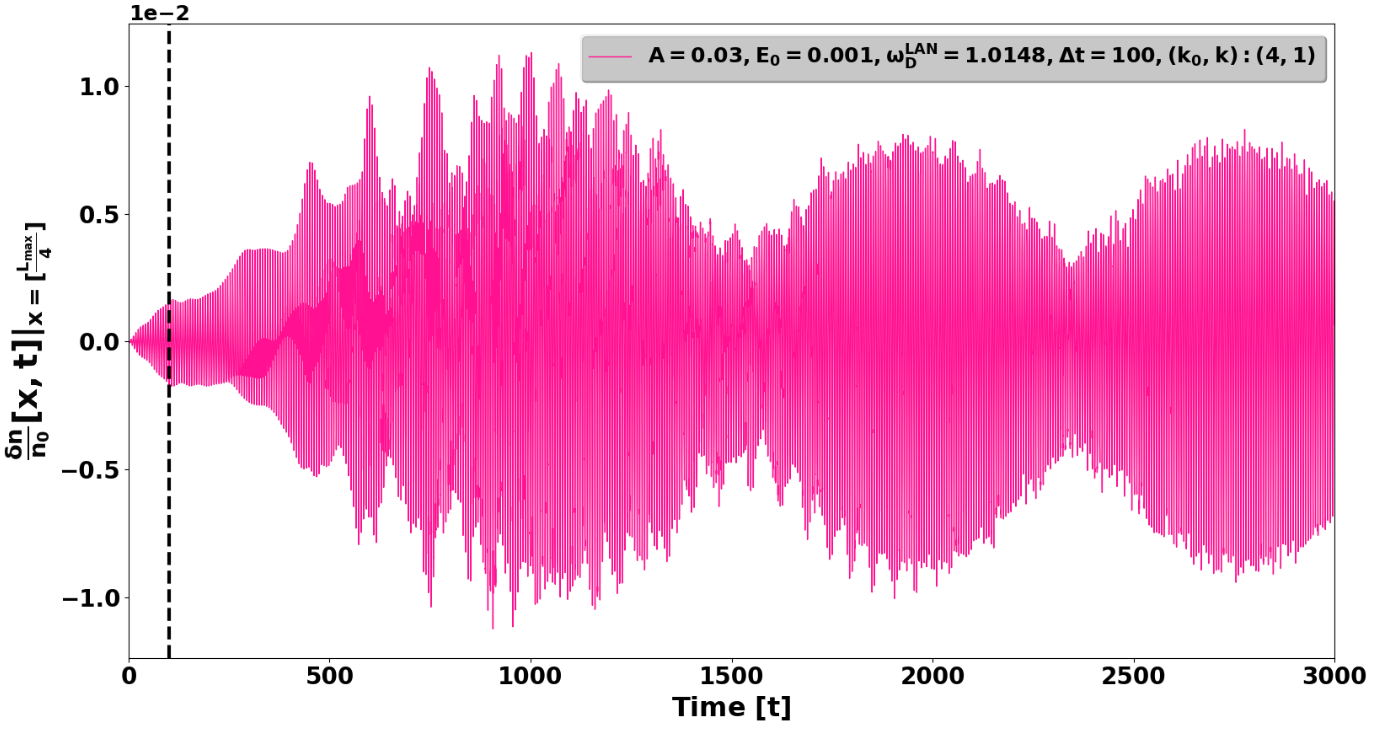}}
\caption{ Temporal variation of excess density fraction (EDF)$:(\delta n/n_{0})$ at $x=L_{max}/4$ where $L_{max}=20\pi=62.8318$ for  $(k_{0},k):(4,1)$, $A=0.03$, $[N_{x} \times N_{v}] = [2048 \times 10000]$, $\omega_{D}^{Lan}=1.0148$, $E_{D}^{0}=0.001$ and $\Delta t=100$ DP case. The dotted black line indicates the time when the external drive is switched off.}
\label{41_EDF_A_0_03}
\end{figure}

 Using excess density fraction (EDF) $: \delta n / n_{0} $ calculations one can quantify the trapped particle fractions which is given as,
 \begin{equation}
 \frac{\delta n}{ n_{0}}(x,t) ~ = ~ \frac{ n(x,t)-n(x,t=0)}{n(x,t=0)}
\label{DIP_EQ_19}
\end{equation}
where $n(x,t)=\int f(x,v,t) dv$ is the density of the electrons i.e first moment of the distribution function. Fig. \ref{41_EDF_A_0_03} illustrate the variation of excess density fraction at $x=L_{max}/4$ where $L_{max}=2\pi/k_{min}$ for $(k_{0},k):(4,1)$, $A=0.03$, $\omega_{D}^{Lan}=1.0148$, $E_{D}^{0}=0.001$ and $\Delta t=100$ DP case. It indicates particle trapping and detrapping modulations due to associated kinetic effects like wave - particle interactions observed in interacting coupled sideband $k/k_{min}=5,~3$ modes.

\begin{figure}
\centerline{\includegraphics[scale=0.28]{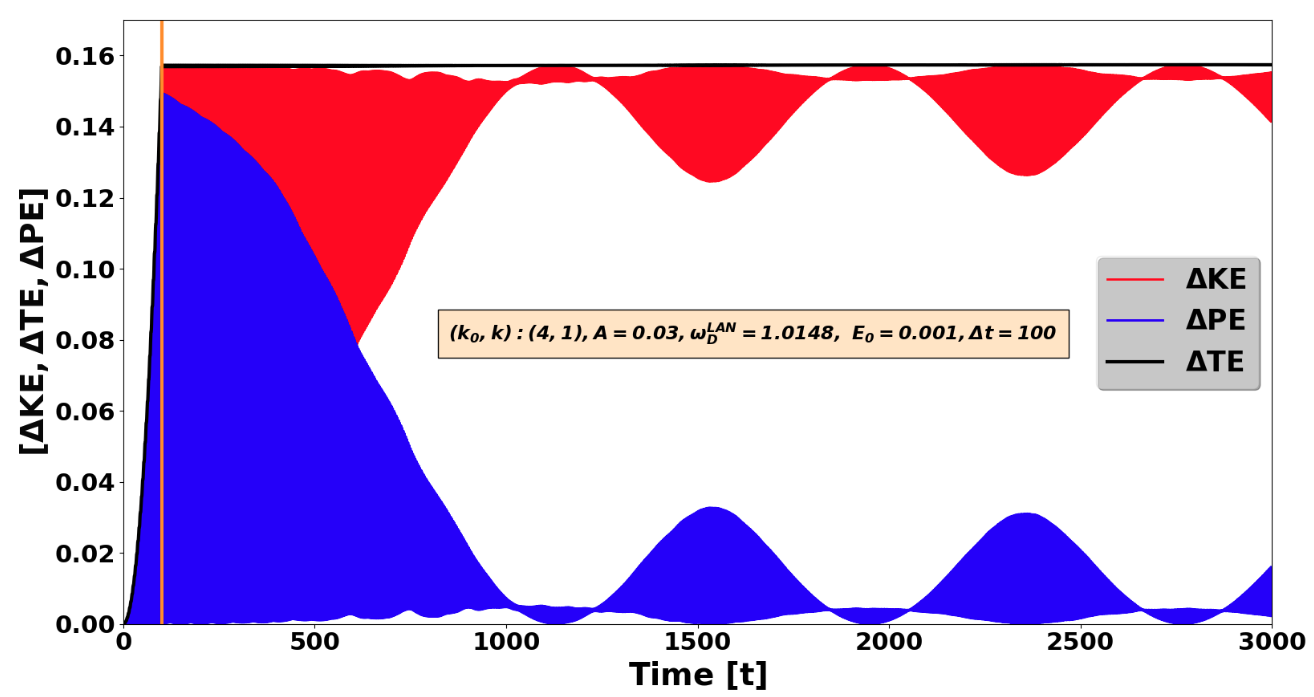}}
\caption{Relative kinetic, potential and total energies $(\Delta KE,~\Delta PE,~\Delta TE)$ variations with respect to time for $(k_{0},k):(4,1)$, $A=0.03$, $[N_{x} \times N_{v}] = [2048 \times 10000]$, $\omega_{D}^{Lan}=1.0148$, $E_{D}^{0}=0.001$ and $\Delta t=100$ DP case. The solid vertical line represents the time when the external electric field drive is switched off.}
\label{41_EC_A=0_03}
\end{figure}

 Numerical consistency of the obtained solutions is demonstrated by plotting the relative difference of kinetic energy $\Delta KE(t)=KE(t)-KE(0)$, potential energy $\Delta PE(t)=PE(t)-PE(0)$ and total energy $\Delta TE(t)=TE(t)-TE(0)$ defined by Eqs. \ref{DIP_EQ_21}, \ref{DIP_EQ_22} with respect to time for $(k_{0},k):(4,1)$, $A=0.03$, $\omega_{D}^{Lan}=1.0148$, $E_{D}^{0}=0.001$ and $\Delta t=100$ DP case in Fig. \ref{41_EC_A=0_03}. It indicates good late time conservation and saturation of relative change in the energies with chosen grid discretization parameters as $N_{x} \times N_{v} = 2048 \times 10000$ for both electrons and ions $(x,v)$ domains after the external electric field drive is switched off.
\begin{equation}
TE(t)=KE(t)+PE(t)
\label{DIP_EQ_20}
\end{equation}
\begin{equation}
KE(t)=\int \int \frac{v^{2}}{2}f(x,v,t)dxdv
\label{DIP_EQ_21}
\end{equation}
\begin{equation}
PE(t)=\int \frac{1}{2}E^{2}(x,t)dx
\label{DIP_EQ_22}
\end{equation}

\begin{figure}
\centerline{\includegraphics[scale=0.27]{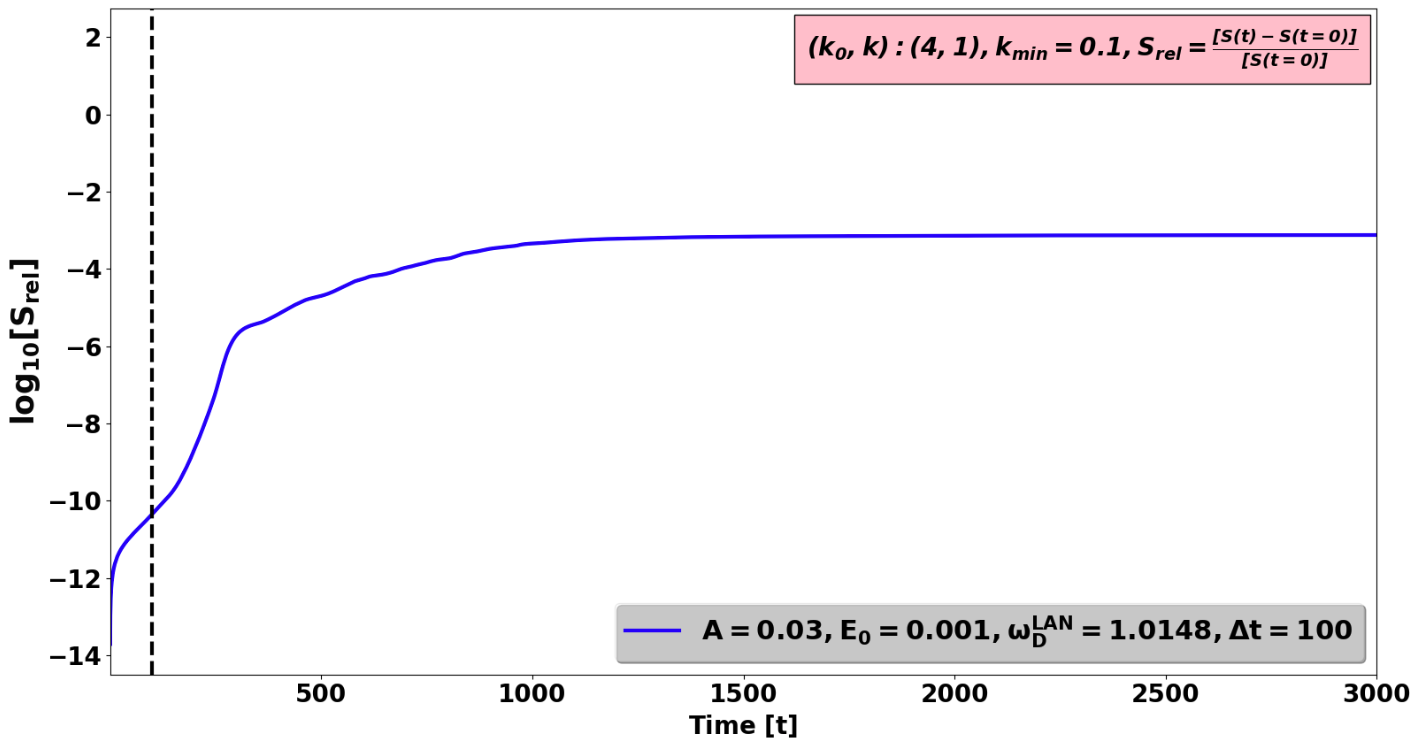}}
\caption{Temporal variation of the relative entropy $S_{rel}$ defined in Eq. \ref{DIP_EQ_23} for $(k_{0},k):(4,1)$, $A=0.03$, $[N_{x} \times N_{v}] = [2048 \times 10000]$, $\omega_{D}^{Lan}=1.0148$, $E_{D}^{0}=0.001$ and $\Delta t=100$ DP case. The dotted vertical line represents the time when the external electric field drive is switched off.}
\label{41_ENTROPY_A=0_03}
\end{figure}

To demonstrate that the obtained solutions are steady state solutions, we have shown the temporal variation of the relative entropy defined by Eqs. \ref{DIP_EQ_23} and \ref{DIP_EQ_24} for $(k_{0},k):(4,1)$, $A=0.03$, $[N_{x} \times N_{v}] = [2048 \times 10000]$, $\omega_{D}^{Lan}=1.0148$, $E_{D}^{0}=0.001$ and $\Delta t=100$ DP case in Fig. \ref{41_ENTROPY_A=0_03}. It act as a measure to observe `information lost' from the system as entropy $S(t)$ has intrinsic tendency for monotonic increase with time due to the ‘ﬁlamentation’ property of the Vlasov - Poisson system \cite{manfredi1997,feix}. The distribution function generates small scale structures in phase space $(x,v)$ due to ﬁlamentation effect, which are dissipated when ﬁlamentation reaches the phase space grid size $[N_{x} \times N_{v}]$ resolutions, leading to saturation in numerical entropy with time as shown in Fig. \ref{41_ENTROPY_A=0_03}. It indicates numerically stable and high quality simulations as the information lost is small with respect to time for chosen grid size resolutions after the external electric field drive is switched off. 
\begin{equation}
S_{rel}(t)=\frac{S(t)-S(t=0)}{S(t=0)}
\label{DIP_EQ_23}
\end{equation} 
\begin{equation}
S(t)=~-\int_{0}^{L_{max}} \int_{-v_{e}^{max}}^{+v_{e}^{max}} f(x,v,t) \log f(x,v,t)dxdv
\label{DIP_EQ_24}
\end{equation} 

  We have demonstrated the coupling of externally driven ``cold'' electron plasma (EP) mode to the plasma bulk via immobile background of inhomogeneous ion analogous to the fluid mode coupling with strong kinetic damping of initial value perturbation (IVP) case presented in Ref. \cite{pandey_2022_KAW}. In the next section, we study the effect of inhomogeneity amplitude on the externally driven plasma cases.
 
\subsection{Inhomogeneous driven plasma case with amplitude $A=0.05$ }
\label{DIP_A=0.05_CASE}

\begin{figure}
\centerline{\includegraphics[scale=0.30]{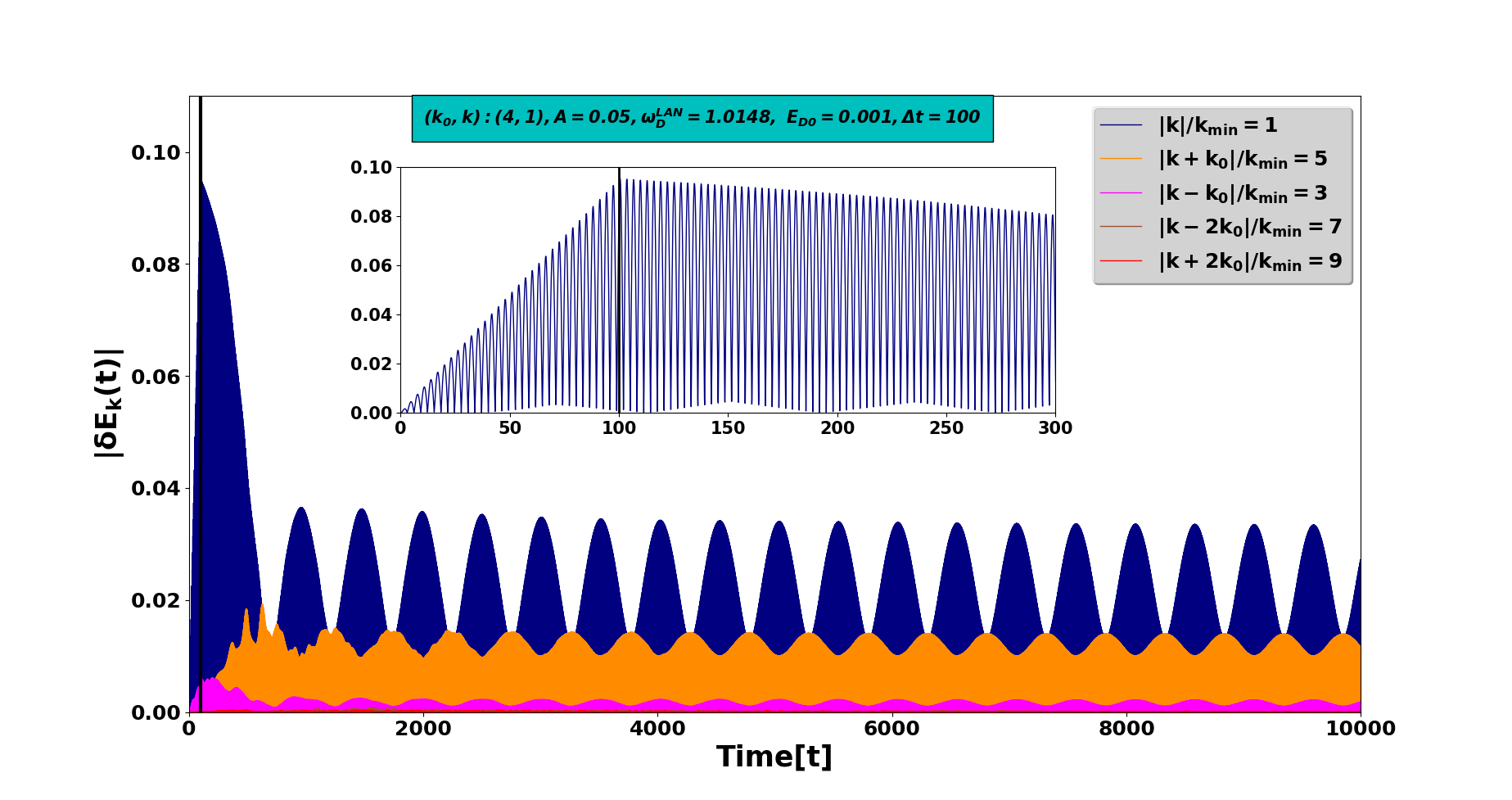}}
\caption{Temporal evolution of perturbed primary as well as coupled interacting sideband Fourier mode amplitude $|\delta E_{k}|$ for externally driven plasma case with $(k_{0},k):(4,1)$, $A=0.03$, $[N_{x} \times N_{v}] = [2048 \times 10000]$, $\omega_{D}^{Lan}=1.0148$, $E_{D}^{0}=0.001$ and $\Delta t=100$. Solid black vertical line indicates the time when the external electric field drive is switched off. Inset plot zooms $|\delta E_{k}|$ for initial time $0$ to $100~\omega_{pe}^{-1}$ illustrating the continuous increase in the electric field amplitude value during the external drive.}
\label{41_LNEK_A=0_05}
\end{figure}

In order to investigate the effect of increase in the background ion inhomogeneous amplitude on the kinetic effects associated with coupled interacting sideband modes as well as wave - wave coupled interactions between primary perturbation and interacting sideband modes more prominently, we present a case with increased background ion inhomogeneity amplitude i.e, $A=0.05$, alongwith identical parameter sets used previously in Sec. \ref{DIP_A=0.03_CASE} for the driven electric field perturbation (DP) case.

Fig. \ref{41_LNEK_A=0_05} demonstrates the variation of Fourier mode signature i.e $|\delta E_{k}|$ of primary and coupled sideband modes with $A=0.05,~(k_{0},k):(4,1),~\omega_{D}^{Lan}=1.0148$, $E_{D}^{0}=0.001$ driven upto $\Delta t=100~\omega_{pe}^{-1}$. It shows damping of the primary perturbation mode initially and late time modulations of the $|\delta E_{k}|$ signature which does not satisfy resonance condition as phase velocity of the mode is around $v_{\phi}^{k}=\omega_{D}^{Lan}/k=10.148$ which falls beyond the velocity domain considered in the problem ($v_{e}^{max}=6.0$). It is important to note that the occurrence of electric field amplitude damping in the primary perturbed mode $k/k_{min}=1$ is mainly due to the wave - wave interaction leading to energy exchange between primary $k/k_{min}=1$ and secondary interacting sideband modes $k/k_{min}=3,~5$ which undergo kinetic damping due to wave - particle resonance interactions (as phase velocities of generated sideband modes $v_{\phi}^{k/k_{min}=3,~5}$ falls in the bulk of the plasma) analogous to the previous case in Sec \ref{DIP_A=0.03_CASE}. Primary perturbed mode $k/k_{min}=1$ does not involve directly in any kinetic phenomenon such as wave - particle interaction as $v_{\phi}^{k/k_{min}=1} >> v_{e}^{max}$. Additionally, from Figs. \ref{41_LNEK_A=0_03}, \ref{41_TAU_FM_VS_A} and \ref{41_LNEK_A=0_05}, one can conclude that with the increase in the ion inhomogeneity amplitude, time required for the energy transfer between primary and sideband modes i.e $\tau_{FM}$ reduces rapidly. However this reduction in $\tau_{FM}$ values is small for larger $A$ values. 

\begin{figure}
\centerline{\includegraphics[scale=0.30]{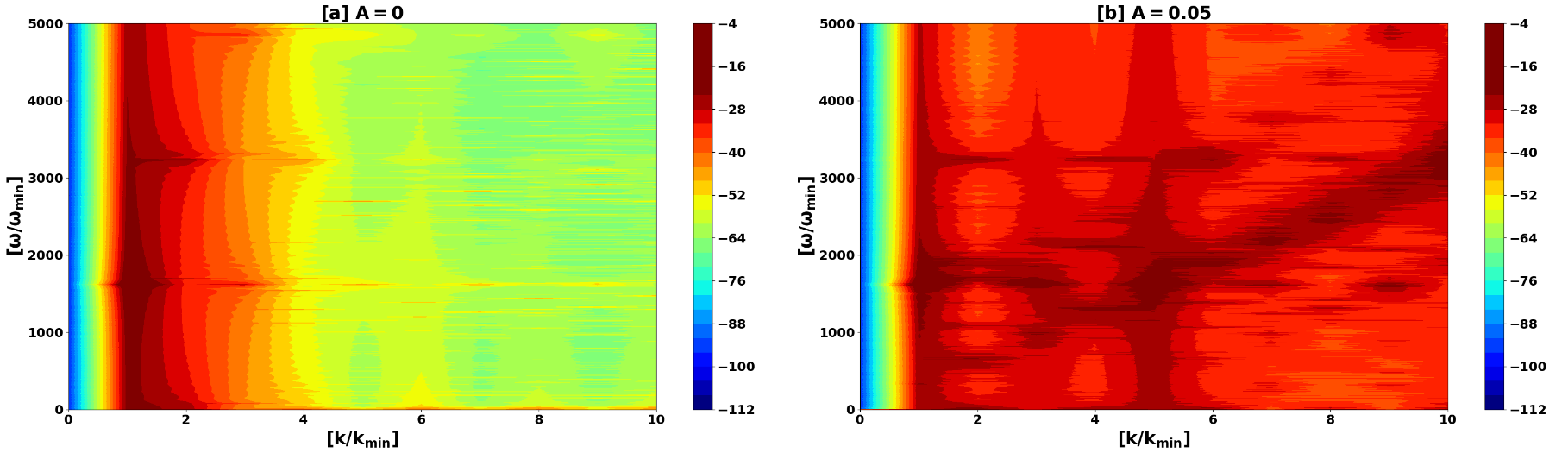}}
\caption{2D power spectrum $(\omega,k)$ contour plot for $(k_{0},k):(4,1)$, $[N_{x} \times N_{v}] = [2048 \times 10000]$, $\omega_{D}^{Lan}=1.0148$, $E_{D}^{0}=0.001$, $\Delta t=100$ driven perturbation with (a) homogeneous $A=0$ and (b) inhomogeneous $A=0.05$ cases. Plot (b) shows the mode coupling signatures between primary perturbation $k/k_{min}=1$ and sideband modes $k/k_{min}=3,~5$ in case of $A \neq 0$. The peak power of the primary mode $k/k_{min}=1$ lies around $\omega / \omega_{min}=1615.1262$ indicating $\omega=1.0148$ as $\omega_{min}=0.00062831$ consistent with 1D Fourier Transform (FFT) analysis.}
\label{41_2DPS_A=0_05}
\end{figure}

\begin{figure}
\centerline{\includegraphics[scale=0.30]{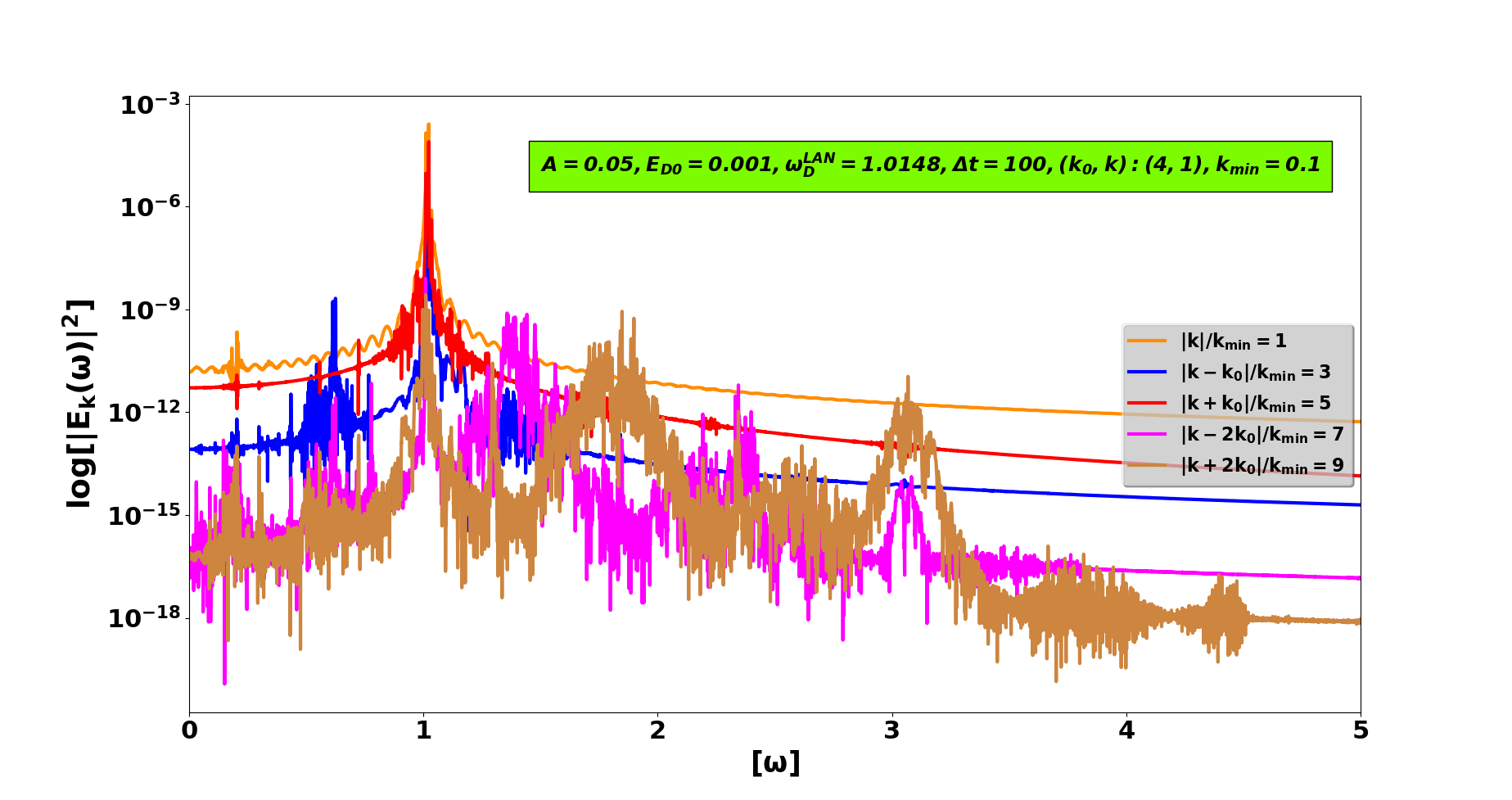}}
\caption{Portrait of power $log(|E_{k}(\omega)|^2 )$ versus frequency $(\omega)$ using 1D Fast Fourier Transform (FFT) analysis for $(k_{0},k):(4,1)$, $N_{x} \times N_{v} = [2048 \times 10000]$, $\omega_{D}^{Lan}=1.0148$, $E_{D}^{0}=0.001$, $\Delta t=100$ DP case with $k_{min}=0.1$. Oscillation frequency of maximum amplitude for primary perturbation mode $k/k_{min}=1$ is $\omega_{k=0.1}=1.02290$ and corresponding frequencies for secondary modes are tabulated in Table \ref{TABLE_2} .}
\label{41_1DFFT_A=0_05}
\end{figure}

 Fig. \ref{41_2DPS_A=0_05} demonstrates the 2D power $(\omega,k)$ spectrum for (a) homogeneous $A=0$ and (b) inhomogeneous $A=0.05$ cases for $(k_{0},k):(4,1)$, $N_{x} \times N_{v} = [2048 \times 10000]$, $k_{min}=0.1$, $\omega_{D}^{Lan}=1.0148$, $E_{D}^{0}=0.001$, $\Delta t=100$ driven perturbation (DP) case. The peak power of the primary perturbation mode $k/k_{min}=1$ in both (a) homogeneous ($A=0$) and (b) inhomogeneous ($A=0.05$) cases lies around $\omega / \omega_{min}=1615.1262$ indicating $\omega=1.0148$ as $\omega_{min}=6.2831 \times 10^{-4}$ consistent with 1D Fourier Transform (FFT) analysis shown in Fig. \ref{41_1DFFT_A=0_05} and tabulated in Table \ref{TABLE_2}. Also, Fig. \ref{41_2DPS_A=0_05} illustrates the mode coupling dynamics between perturbed primary $k/k_{min}=1$ and interacting coupled secondary sideband modes $k/k_{min}=3,~5,~7,~9$ in the presence of non - uniform background of ions with non - zero amplitude. Also, in Fig. \ref{41_2DPS_A=0_05} (a) there is absence of sideband modes and all the power is concentrated in $k/k_{min}=1$ primary perturbation mode, meanwhile in Fig. \ref{41_2DPS_A=0_05} (b), we observe the power transfer and generation of frequencies corresponding to the coupled sideband modes i.e $k/k_{min}=3,~5$ due to mode coupling interaction with the $k/k_{min}=1$ primary perturbed mode.   

In Table \ref{TABLE_2}, we have tabulated the oscillation frequency $(\omega_{k})$ and phase velocity $(v_{\phi}^{k}=\omega_{k}/k)$ corresponding to the primary and secondary interacting sideband modes for $(k_{0},k):(4,1),~E_{D}^{0}=0.001,~A=0.05,~\omega_{D}^{Lan}=1.0148,~\Delta t=100,~k_{min}=0.1$ DP case shown in Fig. \ref{41_1DFFT_A=0_05}. Also, the frequency values obtained through 1D Fast Fourier Transformation (FFT) analysis matches well with the obtained 2D $(\omega,k)$ power spectrum analysis. We can observe that the phase velocities $v_{\phi}^{k}$ values corresponding to sideband $k/k_{min}=3,~5,~7,~9$ modes lies well within $v_{e}^{max}=6.0$ indicating that the resonance locations of these interacting coupled sideband modes lies in the plasma bulk. Hence, energy exchange takes place from primary $k/k_{min}=1$ mode to these sideband modes $k/k_{min}=3,~5,~7,~9$ via mode coupling wave - wave interaction at first, followed by transfer of energy from the coupled sideband modes to the bulk plasma particles via wave - particle interaction i.e Landau damping which finally, causes damping of the externally driven primary $k/k_{min}=1$ mode.

\begin{table}
\caption[Oscillation frequency and Phase Velocity corresponding to the primary and secondary interacting sideband modes ]{Oscillation frequency $(\omega_{k})$ and Phase Velocity $(v_{\phi}^{k}=\omega_{k}/k)$ corresponding to the primary and secondary interacting sideband modes for $(k_{0},k):(4,1),~E_{D}^{0}=0.001,~A=0.05,~\omega_{D}^{Lan}=1.0148,~\Delta t=100,~k_{min}=0.1$ case.}   
\centering                         
\begin{tabular}{c c c}           
\hline\hline                        
Mode No. & $[\omega_{k}^{max}]$ & $[v_{\phi}^{k}=\omega_{k}^{max}/k]$  \\ [1.0ex]    
\hline          
$k=0.1$ & 1.02290 & 10.2290 \\       
$|k-k_{0}|=0.3$ & 1.02290 & 3.4096 \\
$|k+k_{0}|=0.5$ & 1.02290 & 2.0458 \\
$|k-2k_{0}|=0.7$ & 1.00984 & 1.4426 \\
$|k+2k_{0}|=0.9$ & 1.01034 & 1.1226 \\ [1ex]
\hline\hline                               
\end{tabular}
\label{TABLE_2}
\end{table}

\begin{figure}
\centerline{\includegraphics[scale=0.30]{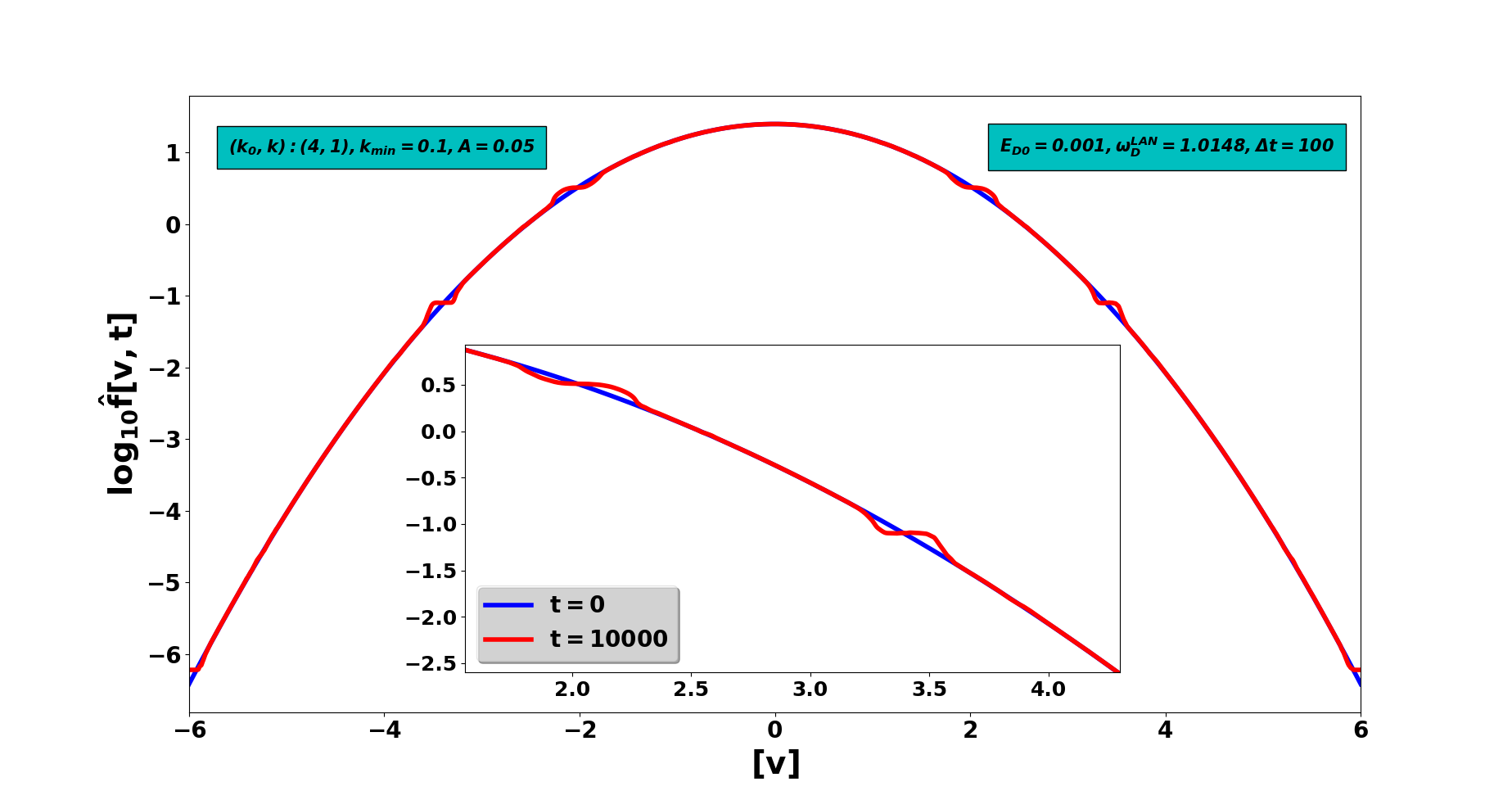}}
\caption{Variation of spatially averaged distribution function $\widehat{f}(v,t)$ with respect to velocity $(v)$ at times $t=0,~10000 ~\omega_{pe}^{-1}$ for $(k_{0},k):(4,1)$, $A=0.05$, $\omega_{D}^{Lan}=1.0148$, $E_{D}^{0}=0.001$ and $\Delta t=100$ DP case. In the inset plot, we can observe bump around phase velocity locations $v_{\phi}^{k}=3.4096,~2.0458$ which corresponds to the coupled $k/k_{min}=3,~5$ sideband modes respectively. Phase velocities corresponding each interacting mode is tabulated in Table \ref{TABLE_2}}
\label{41_DFE_A=0_05}
\end{figure}

\begin{figure}
\centerline{\includegraphics[scale=0.30]{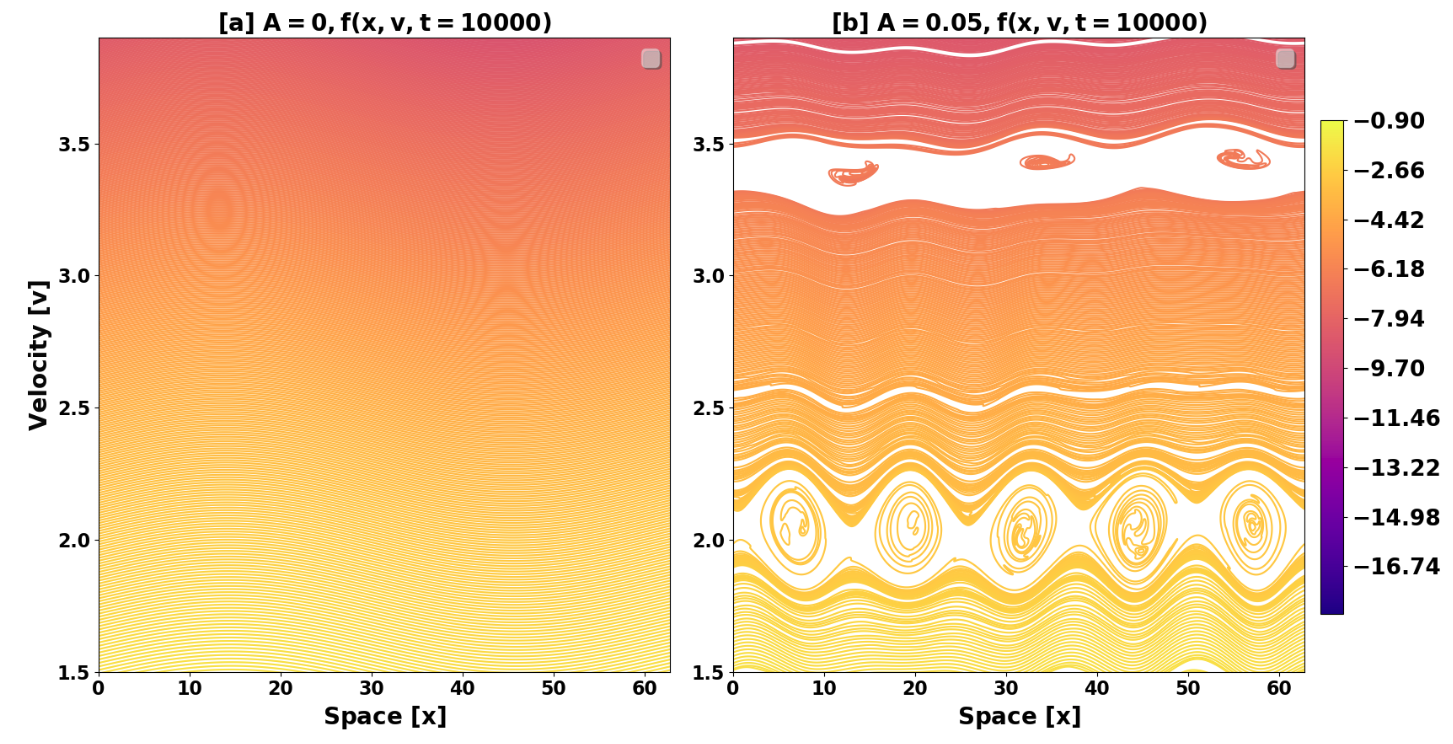}}
\caption{Phase space portrait of electron distribution function $f_{e}(x,v,t)$ at late time $t=10000~\omega_{pe}^{-1}$ for (a) homogeneous $A=0$ and (b) inhomogeneous $A=0.05$ with $(k_{0},k):(4,1)$, $\omega_{D}^{Lan}=1.0148$, $E_{D}^{0}=0.001$ and $\Delta t=100$ DP case. In (b) we can observe the formation of phase space vortices due to particle trapping around resonance locations $v_{\phi}^{k}=3.4096,~2.0458$ corresponding to $k/k_{min}=3,~5$ respectively as tabulated in Table \ref{TABLE_2}. }
\label{41_CP_A=0_A=0_05}
\end{figure}

\begin{figure}
\centerline{\includegraphics[scale=0.32]{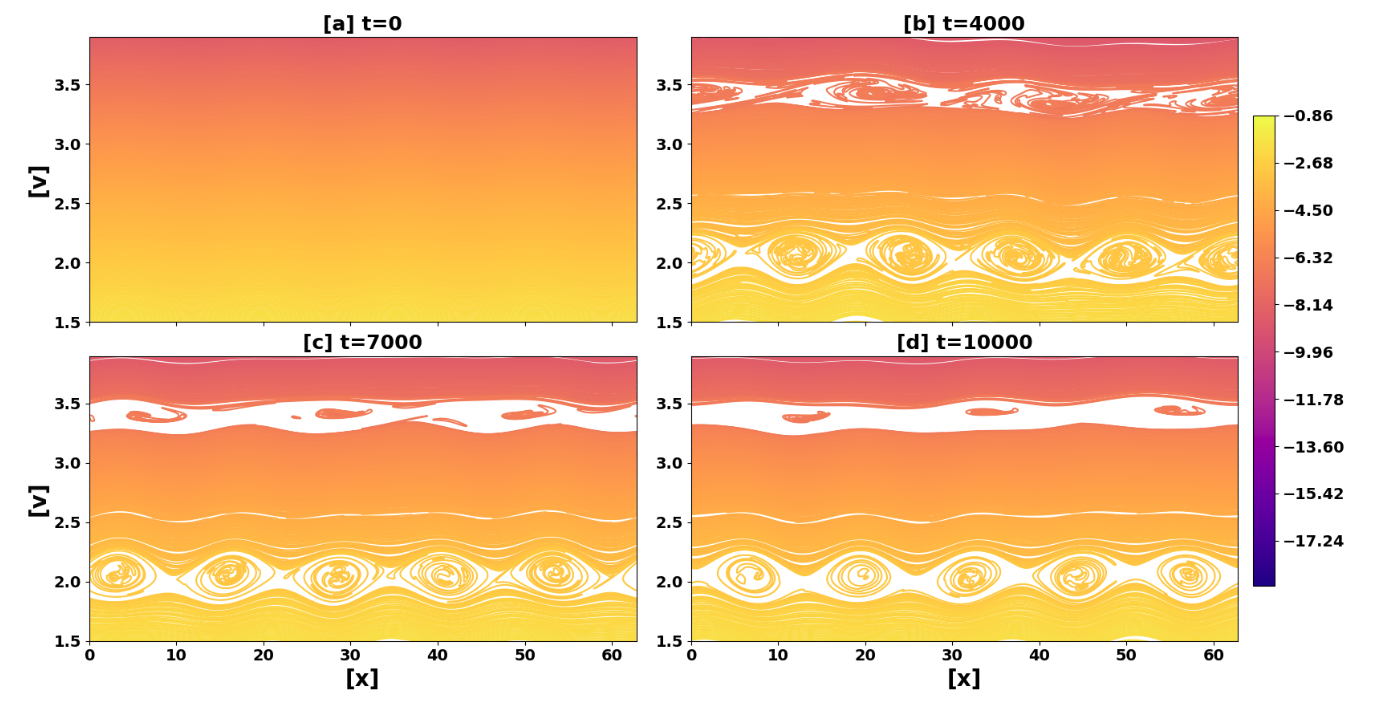}}
\caption{Phase space portrait of electron distribution function $f_{e}(x,v,t)$ at different time (a) $t=0~\omega_{pe}^{-1}$, (b) $t=4000~\omega_{pe}^{-1}$, (c) $t=7000~\omega_{pe}^{-1}$ and (d) $t=10000~\omega_{pe}^{-1}$ for $A=0.05,~(k_{0},k):(4,1)$, $\omega_{D}^{Lan}=1.0148$, $E_{D}^{0}=0.001$ and $\Delta t=100$ DP case. One can observe temporal evolution of vortex structures around resonance locations $v_{\phi}^{k}=3.4096,~2.0458$ corresponding to $k/k_{min}=3,~5$ sideband modes respectively as tabulated in Table \ref{TABLE_2}.}
\label{41_CP_A=0_05}
\end{figure}

Fig. \ref{41_DFE_A=0_05} shows the logarithmic change of spatially averaged distribution function $\widehat{f}(v,t)$ defined by Eq. \ref{DIP_EQ_18} with respect to velocity $(v)$ at initial and final times i.e $t=0,~10000 ~\omega_{pe}^{-1}$ for $(k_{0},k):(4,1)$ with $E_{D}^{0}=0.001,~A=0.05,~k_{min}=0.1,\omega_{D}^{Lan}=1.0148,~\Delta t=100$ DP case. Inset plot shows the zoomed variation of $\widehat{f}(v,t)$ around the resonance velocity $v_{\phi}^{k}$ locations. Formation of plateau or hump region i.e ``local flattening'' around the resonance location $v_{\phi}^{k}=3.4096,~2.0458$ corresponding to the coupled sideband modes $k/k_{min}=3,~5$ indicating the resonating wave - particle energy exchange interaction leading to the particle trapping structures. It is also important to note that we have observed plateau formation around sideband like $k/k_{min}=2$ in the IVP case [Sec. 3.4 of Ref. \cite{pandey_2022_KAW}], whereas despite the increasing the ion inhomogeneity amplitude value which is a major factor in mode coupling dynamics between interacting modes \cite{sanjeev2021,Pandey_2021_TPI_1,Pandey_2021_TPI_2,pandey_2022_KAW}, we are unable to engage the $k/k_{min}=2$ sideband in the DP case [See Figs. \ref{41_CP_A=0_A=0_05} and \ref{41_CP_A=0_05}].


\begin{figure}
\centerline{\includegraphics[scale=0.30]{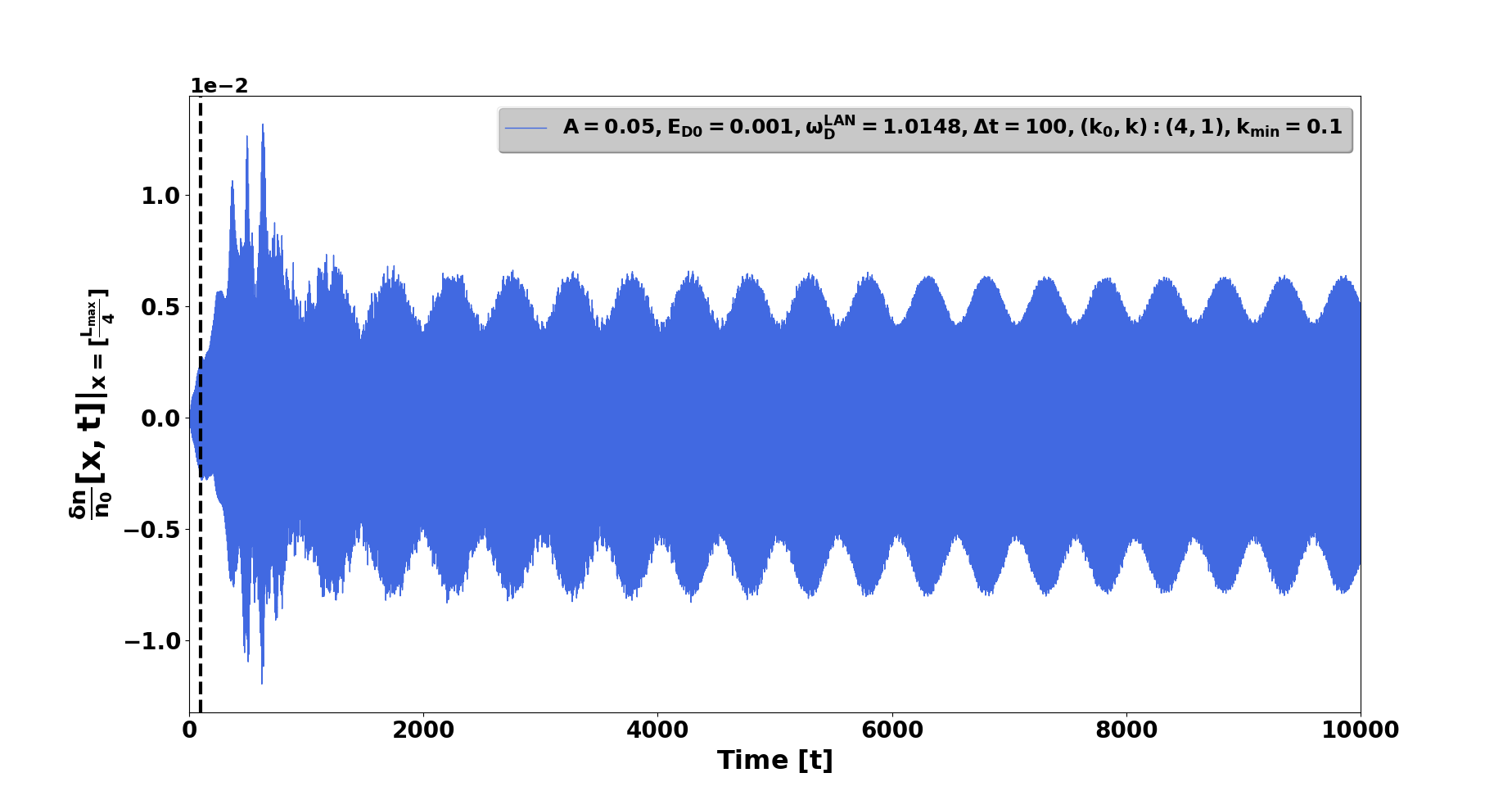}}
\caption{Variation of excess density fraction (EDF)$:(\delta n/n_{0})$ at $x=L_{max}/4$ where $L_{max}=20\pi=62.8318$ with respect to time for  $(k_{0},k):(4,1)$, $A=0.05$, $[N_{x} \times N_{v}] = [2048 \times 10000]$, $\omega_{D}^{Lan}=1.0148$, $E_{D}^{0}=0.001$ and $\Delta t=100$ DP case. The dotted black line indicates the time when the external drive is switched off.}
\label{41_EDF_A=0_05}
\end{figure}

In Fig. \ref{41_CP_A=0_A=0_05}, phase space portrait of the electron distribution function $f_{e}(x,v)$ at late time $t=10000~\omega_{pe}^{-1}$ is shown for (a) homogeneous $A=0$ and (b) $A=0.05$ inhomogeneous plasma cases with $(k_{0},k):(4,1)$, $A=0.05$, $\omega_{D}^{Lan}=1.0148$, $E_{D}^{0}=0.001$ and $\Delta t=100$ DP case. It demonstrates that for the externally driven plasma systems in the absence of ion inhomogeneity i.e for homogeneous $A=0$ case, one can observe absence of phase space vortex structures [Fig. \ref{41_CP_A=0_A=0_05} (a)], due to lack of wave - wave mode coupling and wave - particle interactions, whereas, in the presence of background ion inhomogeneity i.e $A \neq 0$ cases, energy exchange via wave - wave mode coupling interactions between primary perturbation $k/k_{min}=1$ mode and the coupled sideband $k/k_{min}=3,~5$ modes takes place which leads to the formation of phase space vortices as a result of resonant energy exchange interactions between particles and sideband modes at phase velocity $v_{\phi}^{k}$ locations as tabulated in Table. \ref{TABLE_2}. Fig. \ref{41_CP_A=0_05} shows the phase space portrait of electron distribution function $f_{e}(x,v)$ at different time (a) $t=0~\omega_{pe}^{-1}$, (b) $t=4000~\omega_{pe}^{-1}$, (c) $t=7000~\omega_{pe}^{-1}$ and (d) $t=10000~\omega_{pe}^{-1}$ for $A=0.05,~(k_{0},k):(4,1)$, $\omega_{D}^{Lan}=1.0148$, $E_{D}^{0}=0.001$ and $\Delta t=100$ DP case, illustrating the temporal evolution of the electron distribution function in phase space $(x,v)$. In Fig. \ref{41_CP_A=0_05} [(b), (c) and (d)], vortex structure streams with vortivity 3 and 5 belongs to coupled sideband modes $k/k_{min}=3,~5$ respectively indicating wave - particle resonance interactions at $v_{\phi}^{k}=3.4096,~2.0458$ locations [Table \ref{TABLE_2}]. Also, it signifies the non-participation of the sideband modes other than $k/k_{min}=3,~5$ in DP cases which are seen in the IVP cases \cite{pandey_2022_KAW}.

\begin{figure}
\centerline{\includegraphics[scale=0.30]{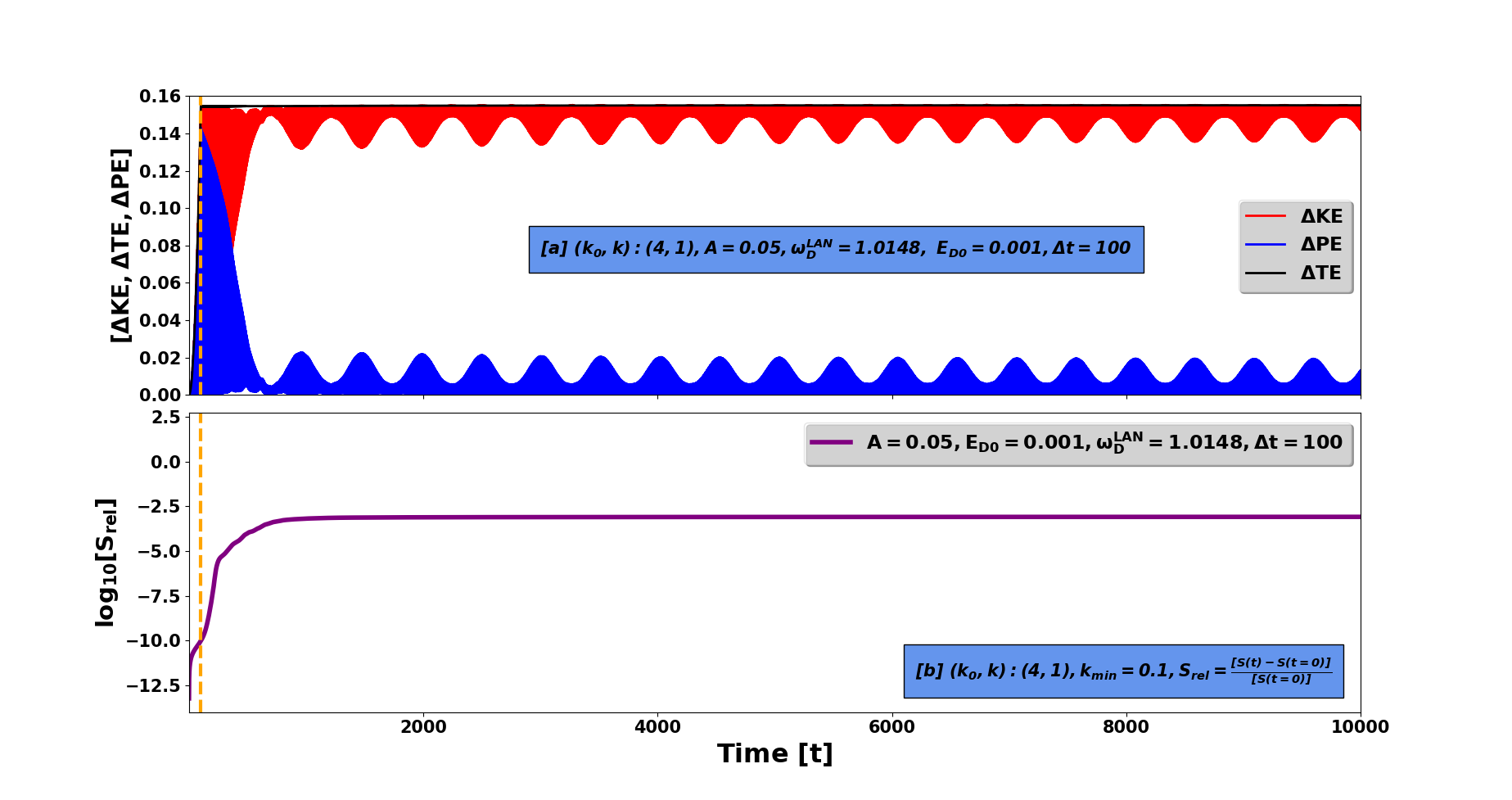}}
\caption{Temporal variation of the (a) relative kinetic, potential and total energies $(\Delta KE,~\Delta PE,~\Delta TE)$ defined in Eqs. \ref{DIP_EQ_21}, \ref{DIP_EQ_22} respectively, and (b) relative entropy $S_{rel}$ defined in Eq. \ref{DIP_EQ_23} for $(k_{0},k):(4,1)$, $A=0.05$, $k_{min}=1$, $[N_{x} \times N_{v}] = [2048 \times 10000]$, $\omega_{D}^{Lan}=1.0148$, $E_{D}^{0}=0.001$ and $\Delta t=100$ DP case. The dotted vertical line at $t=100~\omega_{pe^{-1}}$ represents the time when the external electric field drive is switched off.}
\label{41_EC_A=0_05}
\end{figure}

We show the excess density fraction (EDF) $: \delta n / n_{0} $ variations in Fig. \ref{41_EDF_A=0_05} which demonstrates the variation of EDF at $x=L_{max}/4$ where $L_{max}=2\pi/k_{min}$ for $(k_{0},k):(4,1)$, $A=0.05$, $\omega_{D}^{Lan}=1.0148$, $E_{D}^{0}=0.001$, $\Delta t=100$ DP case indicating the particle trapping and detrapping phenomenon due to resonant interactions between particles and coupled sideband modes. From Figs. \ref{41_EDF_A_0_03} and \ref{41_EDF_A=0_05}, it is obvious that increase in the ion inhomogeneity amplitude $A$ leads to the frequent particle trapping and detrapping modulations. Fig. \ref{41_EC_A=0_05} shows the (a) relative kinetic, potential and total energies $(\Delta KE,~\Delta PE,~\Delta TE)$ defined in Eqs. \ref{DIP_EQ_21}, \ref{DIP_EQ_22} respectively, and (b) relative entropy $S_{rel}$ defined in Eq. \ref{DIP_EQ_23} for $(k_{0},k):(4,1)$, $A=0.05$, $k_{min}=1$, $\omega_{D}^{Lan}=1.0148$, $E_{D}^{0}=0.001$ and $\Delta t=100$ DP case with grid resolution $N_{x} \times N_{v} = [2048 \times 10000]$ in $(x,v)$ domains for both the electrons and ions. The dotted vertical line at $t=100~\omega_{pe}^{-1}$ represents the time when the external electric field drive is switched off. As already mentioned in sec. \ref{DIP_A=0.03_CASE}, time invariance shown in these plots for energy and entropy calculations when the external electric field drive is switched off, act as an indicator of numerically stable, high resolution, steady state simulations with the chosen grid discretizations $[N_{x} \times N_{v}]$ in space and velocity $(x,v)$ domains.

\begin{figure}
\centerline{\includegraphics[scale=0.32]{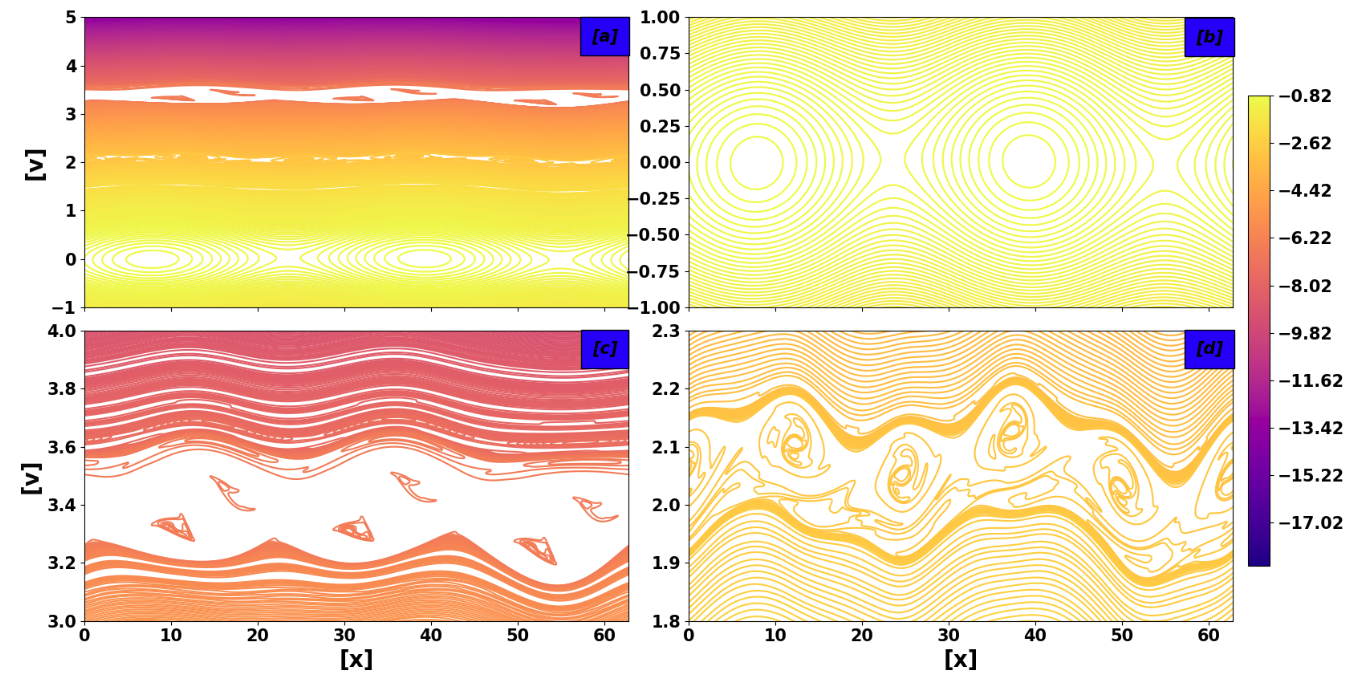}}
\caption{Phase space portrait of electron distribution function $f_{e}(x,v,t)$ at time $t=10000~\omega_{pe}^{-1}$ for $A=0.10,~(k_{0},k):(2,1)$, $\omega_{D}^{Lan}=1.0148$, $E_{D}^{0}=0.001$, $\Delta t=100$ DP case. In (b), vortex chain at $v=0$ is due to chosen equilibrium inhomogeneity scale $k_{0}/k_{min}=2$. In (c), we can observe two separate streams of phase space vortex structures corresponding to $k/k_{min}=3$ mode. In (d), phase space vortex structure stream corresponding to $k/k_{min}=5$ sideband mode is shown.}
\label{21_CP_A=0_10}
\end{figure}

 To illustrate the effect of inhomogeneity scale $k_{0}$ and perturbation scale $k$ dynamics in spatially non - uniform driven plasma systems, with rest of the parameter sets identical to the DP cases, we simulated a case where $k_{0}/k_{min}=2$ instead of $k_{0}/k_{min}=4$ along-with inhomogeneity amplitude $A=0.10$. We only demonstrate Fig. \ref{21_CP_A=0_10} [(a), (b), (c) and (d)] which shows phase space portrait of electron distribution function $f_{e}(x,v,t)$ at time $t=10000~\omega_{pe}^{-1}$ for $A=0.10,~(k_{0},k):(2,1)$, $\omega_{D}^{Lan}=1.0148$, $E_{D}^{0}=0.001$, $\Delta t=100$ DP case. In Fig. \ref{21_CP_A=0_10} (b), phase space vorteex chain at $v=0$ is due to choosen equilibrium inhomogeneity $k_{0}/k_{min}=2$ with two periodic vortices. On detailed inspection of this case, we draw the conclusion that physics behind the damping and associated energy exchange mechanisms for this out of phase primary driven perturbation $k/k_{min}=1$ mode is analogous to the previous $A \neq 0$ driven inhomogeneous cases. Meanwhile, the sidebands created for this case will govern according to $|k \pm Nk_{0}|$ \cite{kaw1973,sanjeev2021} where $N$ is the mode coupling parameter defined in Sec. \ref{DIP_A=0.03_CASE}. Also, in Fig. \ref{21_CP_A=0_10} (c), we see the existence of two different vortex structure streams belonging to the same sideband $k/k_{min}=3$ which is quite different from the previous $A \neq 0$ DP plasma cases but analogous to some extent to the IVP case shown in Ref. \cite{pandey_2022_KAW}.    


\section{Discussion and Conclusion}
\label{Discussion_conclusion}

In this work, we have presented an investigation on the interaction of a primary ``cold" ($v_{\phi} \gg v_{thermal}$) electron plasma (EP) mode coupling as well as energy exchange dynamics created by low amplitude, constant frequency (Langmuir frequency $\omega_{D}^{Lan}$) external electric field drive, in the presence of stationary background of inhomogeneous ions. We have also compared the solutions obtained from two different perturbation approaches i.e initial value perturbation (IVP) approach and driven perturbation (DP) approach for the identical simulation parameters. The results are summarized as follows $: \longrightarrow$\\

\begin{itemize}
  \item We have demonstrated that externally driven electron plasma mode with phase velocity $v_{\phi} \gg v_{thermal} : v_{\phi} \gg v_{e}^{max}$ can be coupled to the plasma bulk via stationary background of non - uniform ions through wave - wave and wave - particle interaction phenomenon analogous to initial value perturbation (IVP) cases.

  \item In both IVP and driven perturbation (DP) cases, from simulation data we infer that time taken by the modes to reach the first minimum of the field signature i.e $\tau_{FM}$ is inversely proportional to ion inhomogeneity amplitude [Figs. \ref{41_LNEK_A=0_03}, \ref{41_TAU_FM_VS_A} and \ref{41_LNEK_A=0_05}]. 

  \item In Fig. 20 of Ref. \cite{pandey_2022_KAW}, we found two vortex structures in phase space, around $v_{\phi} = 5.290$ due to $k/k_{min} = 2$ sideband mode, which is absent in the DP case as shown in Fig. \ref{41_CP_A=0_03}. Also, it is important to note that with the increased ion inhomogeneity amplitude DP case with $A=0.05$, we are unable to engage $k/k_{min}=2$ sideband mode. 

  \item From Fig. \ref{41_CP_A=0_A=0_05}, by illustrating the 2D $(\omega.k)$ power spectrum for $A=0.05$ DP case, we clearly demonstrate the distinct role of wave-wave mode coupling dynamics between driven primary perturbation $k/k_{min}=1$ mode and interacting coupled sideband $k/k_{min}=3,~5,~7,~9$ modes in the inhomogeneous plasma cases which is absent in the homogeneous plasma $A=0$ DP cases.  

  \item As we understand, following is the energy exchange mechanisms occurring between driven $k/k_{min}=1$ wave, sidebands $k/k_{min}=3,~5,~7,~9$ and particles which leads to the damping of out of phase primary perturbation mode with $v_{\phi} \gg v_{thermal}$ : $v_{\phi} \gg v_{e}^{max}$  $\longrightarrow$
    \begin{itemize}
    
        \item At first, energy exchange between primary perturbed $k/k_{min}=1$ and sidebands $k/k_{min}=3,~5,~7,~9$ occurs via wave - wave mode coupling interactions.

        \item Next, phase velocities corresponding to the sideband modes $k/k_{min}=3,~5,~7,~9$ falls inside the plasma bulk i.e $v_{\phi}^{k} \sim v_{thermal}$ leading to energy exchange between sidebands and particles through resonance wave - particle interactions (Landau damping).

        \item Finally, energy from the externally driven primary perturbation $k/k_{min}=1$ mode is transferred into the plasma.
        
    \end{itemize}
  
  \item In the present work, kinetic phenomena such as Landau damping, wave -wave mode coupling energy exchange interactions, formation of phase space vortex structure or BGK waves etc, are demonstrated for the externally driven plasma system in the presence of non-uniform background of stationary ions.  
 
\end{itemize}

Our findings brings out several key features related to externally driven systems in the presence of background spatial ion non-uniformity which may be crucial in understanding the phenomena such as wave-wave mode coupling dynamics and collisionless plasma turbulence etc, relevant to the plasmas in laboratory as well as astrophysical settings. Our future investigations will be based on the study of plasma systems consisting of several modes interacting due to non-linearity and mode coupling in the presence of a spectrum of ion waves excited with various types of external drivers.

\section*{Acknowledgments}
All the computational results of this paper were obtained using the ANTYA HPC Linux cluster at Institute for Plasma Research (IPR) Gandhinagar, India. The authors would like to thank the Data Center staff at IPR.


\section*{Data availability statement}
The data that support the findings of this study are available upon reasonable request from the authors.

\section*{References}
\bibliography{iopart-num}

\providecommand{\newblock}{}
\begin{thebibliography}{10}
\expandafter\ifx\csname url\endcsname\relax
  \def\url#1{{\tt #1}}\fi
\expandafter\ifx\csname urlprefix\endcsname\relax\def\urlprefix{URL }\fi
\providecommand{\eprint}[2][]{\url{#2}}

\bibitem{tonks_langmuir1929}
Tonks L and Langmuir I 1929 {\em Phys. Rev.\/} {\bf 33}(2) 195--210
  \urlprefix\url{https://link.aps.org/doi/10.1103/PhysRev.33.195}

\bibitem{bohm1949}
Bohm D and Gross E~P 1949 {\em Phys. Rev.\/} {\bf 75}(12) 1851--1864
  \urlprefix\url{https://link.aps.org/doi/10.1103/PhysRev.75.1851}

\bibitem{landau1946}
Haar D 2013 {\em Collected Papers of L.D. Landau\/} (Elsevier Science) ISBN
  9781483152707
  \urlprefix\url{https://books.google.co.in/books?id=epc4BQAAQBAJ}

\bibitem{kampen1955}
{Van Kampen} N 1955 {\em Physica\/} {\bf 21} 949--963 ISSN 0031-8914
  \urlprefix\url{https://www.sciencedirect.com/science/article/pii/S0031891455930688}

\bibitem{bgk1957}
Bernstein I~B, Greene J~M and Kruskal M~D 1957 {\em Phys. Rev.\/} {\bf 108}(3)
  546--550 \urlprefix\url{https://link.aps.org/doi/10.1103/PhysRev.108.546}

\bibitem{dawson1959}
Dawson J~M 1959 {\em Phys. Rev.\/} {\bf 113}(2) 383--387
  \urlprefix\url{https://link.aps.org/doi/10.1103/PhysRev.113.383}

\bibitem{oneil1965}
O'Neil T 1965 {\em The Physics of Fluids\/} {\bf 8} 2255--2262
  (\textit{Preprint}
  \eprint{https://aip.scitation.org/doi/pdf/10.1063/1.1761193})
  \urlprefix\url{https://aip.scitation.org/doi/abs/10.1063/1.1761193}

\bibitem{kds1969}
Kruer W~L, Dawson J~M and Sudan R~N 1969 {\em Phys. Rev. Lett.\/} {\bf 23}(15)
  838--841 \urlprefix\url{https://link.aps.org/doi/10.1103/PhysRevLett.23.838}

\bibitem{goldman1970}
Goldman M~V 1970 {\em The Physics of Fluids\/} {\bf 13} 1281--1289
  (\textit{Preprint}
  \eprint{https://aip.scitation.org/doi/pdf/10.1063/1.1693061})
  \urlprefix\url{https://aip.scitation.org/doi/abs/10.1063/1.1693061}

\bibitem{gary_tokar_1985}
Gary S~P and Tokar R~L 1985 {\em The Physics of Fluids\/} {\bf 28} 2439--2441
  ISSN 0031-9171 (\textit{Preprint}
  \eprint{https://pubs.aip.org/aip/pfl/article-pdf/28/8/2439/12595671/2439\_1\_online.pdf})
  \urlprefix\url{https://doi.org/10.1063/1.865250}

\bibitem{valentini_prl_2010}
Valentini F, Califano F and Veltri P 2010 {\em Phys. Rev. Lett.\/} {\bf
  104}(20) 205002
  \urlprefix\url{https://link.aps.org/doi/10.1103/PhysRevLett.104.205002}

\bibitem{valentini_prl_2011}
Valentini F, Califano F, Perrone D, Pegoraro F and Veltri P 2011 {\em Phys.
  Rev. Lett.\/} {\bf 106}(16) 165002
  \urlprefix\url{https://link.aps.org/doi/10.1103/PhysRevLett.106.165002}

\bibitem{Valentini_2011}
Valentini F, Califano F, Perrone D, Pegoraro F and Veltri P 2011 {\em Plasma
  Physics and Controlled Fusion\/} {\bf 53} 105017
  \urlprefix\url{https://dx.doi.org/10.1088/0741-3335/53/10/105017}

\bibitem{valentini_2012}
Valentini F, Perrone D, Califano F, Pegoraro F, Veltri P, Morrison P~J and
  O'Neil T~M 2012 {\em Physics of Plasmas\/} {\bf 19} 092103 ISSN 1070-664X
  (\textit{Preprint}
  \eprint{https://pubs.aip.org/aip/pop/article-pdf/doi/10.1063/1.4751440/16126006/092103\_1\_online.pdf})
  \urlprefix\url{https://doi.org/10.1063/1.4751440}

\bibitem{pallavi_2016}
Trivedi P and Ganesh R 2016 {\em Physics of Plasmas\/} {\bf 23} 062112 ISSN
  1070-664X (\textit{Preprint}
  \eprint{https://pubs.aip.org/aip/pop/article-pdf/doi/10.1063/1.4953603/15903366/062112\_1\_online.pdf})
  \urlprefix\url{https://doi.org/10.1063/1.4953603}

\bibitem{pallavi_2017}
Trivedi P and Ganesh R 2017 {\em Physics of Plasmas\/} {\bf 24} 032107 ISSN
  1070-664X (\textit{Preprint}
  \eprint{https://pubs.aip.org/aip/pop/article-pdf/doi/10.1063/1.4978560/15934374/032107\_1\_online.pdf})
  \urlprefix\url{https://doi.org/10.1063/1.4978560}

\bibitem{pallavithesis}
Trivedi P 2019 {\em Driven Phase Space Structures In A 1D Vlasov-Poisson
  Plasma\/} Ph.D. thesis Institute for Plasma Research

\bibitem{pallavi_2018}
Trivedi P and Ganesh R 2018 {\em Physics of Plasmas\/} {\bf 25} 112102 ISSN
  1070-664X (\textit{Preprint}
  \eprint{https://pubs.aip.org/aip/pop/article-pdf/doi/10.1063/1.5052494/15953062/112102\_1\_online.pdf})
  \urlprefix\url{https://doi.org/10.1063/1.5052494}

\bibitem{kruer_kaw_1970}
Kruer W~L, Kaw P~K, Dawson J~M and Oberman C 1970 {\em Phys. Rev. Lett.\/} {\bf
  24}(18) 987--990
  \urlprefix\url{https://link.aps.org/doi/10.1103/PhysRevLett.24.987}

\bibitem{kruer_prl1970}
Kruer W~L and Dawson J~M 1970 {\em Phys. Rev. Lett.\/} {\bf 25}(17) 1174--1176
  \urlprefix\url{https://link.aps.org/doi/10.1103/PhysRevLett.25.1174}

\bibitem{kruer1972}
Kruer W~L 1972 {\em The Physics of Fluids\/} {\bf 15} 2423--2426
  (\textit{Preprint}
  \eprint{https://aip.scitation.org/doi/pdf/10.1063/1.1693887})
  \urlprefix\url{https://aip.scitation.org/doi/abs/10.1063/1.1693887}

\bibitem{jackson1966}
Jackson E~A and Raether M 1966 {\em The Physics of Fluids\/} {\bf 9} 1257--1259
  (\textit{Preprint}
  \eprint{https://aip.scitation.org/doi/pdf/10.1063/1.1761834})
  \urlprefix\url{https://aip.scitation.org/doi/abs/10.1063/1.1761834}

\bibitem{harding1968}
Harding R~C 1968 {\em The Physics of Fluids\/} {\bf 11} 2233--2240
  (\textit{Preprint}
  \eprint{https://aip.scitation.org/doi/pdf/10.1063/1.1691807})
  \urlprefix\url{https://aip.scitation.org/doi/abs/10.1063/1.1691807}

\bibitem{dorman1970}
Dorman G 1970 {\em Journal of Plasma Physics\/} {\bf 4} 127–142

\bibitem{bertrand_feix_baumann_1971}
Bertrand P, Feix M~R and Baumann G 1971 {\em Journal of Plasma Physics\/} {\bf
  6} 351–366

\bibitem{kaw1973}
Kaw P~K, Lin A~T and Dawson J~M 1973 {\em The Physics of Fluids\/} {\bf 16}
  1967--1975 (\textit{Preprint}
  \eprint{https://aip.scitation.org/doi/pdf/10.1063/1.1694242})
  \urlprefix\url{https://aip.scitation.org/doi/abs/10.1063/1.1694242}

\bibitem{sanjeev2021}
Pandey S~K and Ganesh R 2021 {\em AIP Advances\/} {\bf 11} 025229
  (\textit{Preprint} \eprint{https://doi.org/10.1063/5.0030082})
  \urlprefix\url{https://doi.org/10.1063/5.0030082}

\bibitem{pandey_2022_KAW}
Pandey S~K, Mahapatra J and Ganesh R 2022 {\em Physica Scripta\/} {\bf 97}
  105602 \urlprefix\url{https://dx.doi.org/10.1088/1402-4896/ac90f4}

\bibitem{xu2019}
Xu H, Su F~f, Kong X~m, Sun Y, Jin R~n, Huang G~x and Du S~j 2019 {\em Physics
  of Plasmas\/} {\bf 26} 022112 (\textit{Preprint}
  \eprint{https://doi.org/10.1063/1.5085154})
  \urlprefix\url{https://doi.org/10.1063/1.5085154}

\bibitem{nidhi2021}
Rathee N, Mukherjee A, Trines R~M~G~M and Sengupta S 2021 {\em Physics of
  Plasmas\/} {\bf 28} 012105 (\textit{Preprint}
  \eprint{https://doi.org/10.1063/5.0033658})
  \urlprefix\url{https://doi.org/10.1063/5.0033658}

\bibitem{everett1995}
Everett M~J, Lal A, Clayton C~E, Mori W~B, Johnston T~W and Joshi C 1995 {\em
  Phys. Rev. Lett.\/} {\bf 74}(12) 2236--2239
  \urlprefix\url{https://link.aps.org/doi/10.1103/PhysRevLett.74.2236}

\bibitem{everett1996}
Everett M~J, Lal A, Clayton C~E, Mori W~B, Joshi C and Johnston T~W 1996 {\em
  Physics of Plasmas\/} {\bf 3} 2041--2046 (\textit{Preprint}
  \eprint{https://doi.org/10.1063/1.871678})
  \urlprefix\url{https://doi.org/10.1063/1.871678}

\bibitem{estabrook1981}
Estabrook K 1981 {\em Phys. Rev. Lett.\/} {\bf 47}(19) 1396--1399
  \urlprefix\url{https://link.aps.org/doi/10.1103/PhysRevLett.47.1396}

\bibitem{rosen1972}
Rosen B, Schmidt G and Kruer W~L 1972 {\em The Physics of Fluids\/} {\bf 15}
  2001--2006 (\textit{Preprint}
  \eprint{https://aip.scitation.org/doi/pdf/10.1063/1.1693814})
  \urlprefix\url{https://aip.scitation.org/doi/abs/10.1063/1.1693814}

\bibitem{canosa1976}
Canosa J and Wray A 1976 {\em The Physics of Fluids\/} {\bf 19} 1958--1966
  (\textit{Preprint}
  \eprint{https://aip.scitation.org/doi/pdf/10.1063/1.861413})
  \urlprefix\url{https://aip.scitation.org/doi/abs/10.1063/1.861413}

\bibitem{shoucri1978}
Shoucri M~M 1978 {\em The Physics of Fluids\/} {\bf 21} 1359--1365
  (\textit{Preprint}
  \eprint{https://aip.scitation.org/doi/pdf/10.1063/1.862377})
  \urlprefix\url{https://aip.scitation.org/doi/abs/10.1063/1.862377}

\bibitem{Buchelnikova_1980}
Buchelnikova N~S and Matochkin E~P 1980 {\em Physica Scripta\/} {\bf 22}
  632--636 \urlprefix\url{https://doi.org/10.1088/0031-8949/22/6/014}

\bibitem{Buchelnikova_1981}
Buchelnikova N~S and Matochkin E~P 1981 {\em Physica Scripta\/} {\bf 24}
  566--574 \urlprefix\url{https://doi.org/10.1088/0031-8949/24/3/011}

\bibitem{koch1983}
Koch B~P and Leven R~W 1983 {\em Physica Scripta\/} {\bf 27} 220--224
  \urlprefix\url{https://doi.org/10.1088/0031-8949/27/3/013}

\bibitem{barr1986}
Barr H~C, Boyd T~J~M and Coutts G~A 1986 {\em Phys. Rev. Lett.\/} {\bf 56}(21)
  2256--2259
  \urlprefix\url{https://link.aps.org/doi/10.1103/PhysRevLett.56.2256}

\bibitem{villeneuve1987}
Villeneuve D~M, Baldis H~A and Bernard J~E 1987 {\em Phys. Rev. Lett.\/} {\bf
  59}(14) 1585--1588
  \urlprefix\url{https://link.aps.org/doi/10.1103/PhysRevLett.59.1585}

\bibitem{ghizzo1988}
Ghizzo A, Izrar B, Bertrand P, Fijalkow E, Feix M~R and Shoucri M 1988 {\em The
  Physics of Fluids\/} {\bf 31} 72--82 (\textit{Preprint}
  \eprint{https://aip.scitation.org/doi/pdf/10.1063/1.866579})
  \urlprefix\url{https://aip.scitation.org/doi/abs/10.1063/1.866579}

\bibitem{brunetti2000}
Brunetti M, Califano F and Pegoraro F 2000 {\em Phys. Rev. E\/} {\bf 62}(3)
  4109--4114 \urlprefix\url{https://link.aps.org/doi/10.1103/PhysRevE.62.4109}

\bibitem{manfredi2000}
Manfredi G and Bertrand P 2000 {\em Physics of Plasmas\/} {\bf 7} 2425--2431
  (\textit{Preprint} \eprint{https://doi.org/10.1063/1.874081})
  \urlprefix\url{https://doi.org/10.1063/1.874081}

\bibitem{shoucri_2006}
Shoucri M 2006 {\em Journal of Plasma Physics\/} {\bf 72} 861–864

\bibitem{shukla_2009}
Shukla P~K 2009 {\em Physica Scripta\/} {\bf 80} 038201
  \urlprefix\url{https://doi.org/10.1088/0031-8949/80/03/038201}

\bibitem{yang2020}
Yang T, Feng Q~S, Wang Y~X, Zhou Y~Z, Ban S~S, Zhang S~T, Xie R, Jiang Y, Cao
  L~H, Liu Z~J and Zheng C~Y 2020 {\em Plasma Physics and Controlled Fusion\/}
  {\bf 62} 095009 \urlprefix\url{https://doi.org/10.1088/1361-6587/ab9d68}

\bibitem{Pandey_2021_TPI_2}
Pandey S~K and Ganesh R 2021 {\em Physica Scripta\/} {\bf 96} 125615
  \urlprefix\url{https://doi.org/10.1088/1402-4896/ac25a2}

\bibitem{pottelette1984}
Pottelette R, Illiano J~M, Bauer O~H and Treumann R 1984 {\em Journal of
  Geophysical Research: Space Physics\/} {\bf 89} 2324--2334 (\textit{Preprint}
  \eprint{https://agupubs.onlinelibrary.wiley.com/doi/pdf/10.1029/JA089iA04p02324})
  \urlprefix\url{https://agupubs.onlinelibrary.wiley.com/doi/abs/10.1029/JA089iA04p02324}

\bibitem{guzdar1996}
Guzdar P~N, Chaturvedi P~K, Papadopoulos K, Keskinen M~J and Ossakow S~L 1996
  {\em Journal of Geophysical Research: Space Physics\/} {\bf 101} 2453--2460
  (\textit{Preprint}
  \eprint{https://agupubs.onlinelibrary.wiley.com/doi/pdf/10.1029/95JA02975})
  \urlprefix\url{https://agupubs.onlinelibrary.wiley.com/doi/abs/10.1029/95JA02975}

\bibitem{manfredi1997}
Manfredi G 1997 {\em Phys. Rev. Lett.\/} {\bf 79}(15) 2815--2818
  \urlprefix\url{https://link.aps.org/doi/10.1103/PhysRevLett.79.2815}

\bibitem{raghunathan2013}
Raghunathan M and Ganesh R 2013 {\em Physics of Plasmas\/} {\bf 20} 032106
  (\textit{Preprint} \eprint{https://doi.org/10.1063/1.4794320})
  \urlprefix\url{https://doi.org/10.1063/1.4794320}

\bibitem{Saini2018}
Saini V, Pandey S~K, Trivedi P and Ganesh R 2018 {\em Physics of Plasmas\/}
  {\bf 25} 092107 (\textit{Preprint}
  \eprint{https://doi.org/10.1063/1.5024376})
  \urlprefix\url{https://doi.org/10.1063/1.5024376}

\bibitem{Pandey_2021_TPI_1}
Pandey S~K and Ganesh R 2021 {\em Physica Scripta\/} {\bf 96} 125616
  \urlprefix\url{https://doi.org/10.1088/1402-4896/ac25a1}

\bibitem{sanjeevthesis}
Pandey S~K 2023 {\em Linear and non-linear waves in spatially non-uniform 1D
  Vlasov-Poisson plasmas.\/} Ph.D. thesis Institute for Plasma Research

\bibitem{colella1984}
Colella P and Woodward P~R 1984 {\em Journal of Computational Physics\/} {\bf
  54} 174 -- 201 ISSN 0021-9991
  \urlprefix\url{http://www.sciencedirect.com/science/article/pii/0021999184901438}

\bibitem{cheng1976}
Cheng C and Knorr G 1976 {\em Journal of Computational Physics\/} {\bf 22} 330
  -- 351 ISSN 0021-9991
  \urlprefix\url{http://www.sciencedirect.com/science/article/pii/002199917690053X}

\bibitem{arber_2002}
Arber T and Vann R 2002 {\em Journal of Computational Physics\/} {\bf 180}
  339--357 ISSN 0021-9991
  \urlprefix\url{https://www.sciencedirect.com/science/article/pii/S0021999102970981}

\bibitem{vann_thesis}
Vann R 2002 {\em Characterization of fully nonlinear Berk-Breizman
  phenomenology.\/} Ph.D. thesis University of Warwick.

\bibitem{feix}
Feix M~R, Bertrand P and Ghizzo A 1994 {\em Advances in Kinetic Theory and
  Computing\/} (World Scientific, Singapore. pp. 45–81.)

\end{thebibliography}

\end{document}